\documentclass[aps,prd,twocolumn,showpacs,floats,nofootinbib]{revtex4-1}

\usepackage{graphicx}
\usepackage{amsmath,amsfonts}
\usepackage{dcolumn}
\usepackage{hyperref}

\mathchardef\minus = "002D

\newcommand{\swS}[5][]{{}_{{}_{#2}}S^{#1}_{#3}(#4;#5)}
\newcommand{\scA}[4][]{{}_{{}_{#2}}A^{#1}_{#3}(#4)}

\newcolumntype{f}[1]{D{.}{.}{#1}}

\begin{document}

\title{An unknown branch of the total-transmission modes for the {K}err-geometry}

\author{Gregory B. Cook}\email{cookgb@wfu.edu}
\affiliation{Department of Physics, Wake Forest University,
		 Winston-Salem, North Carolina 27109}
\author{Luke S. Annichiarico}\email{annils14@wfu.edu}
\affiliation{Department of Physics, Wake Forest University,
		 Winston-Salem, North Carolina 27109}
\author{Daniel J. Vickers}\email{vickdj14@wfu.edu}
\affiliation{Department of Physics, Wake Forest University,
		 Winston-Salem, North Carolina 27109}

\date{\today}

\begin{abstract}
  The gravitational modes of the Kerr geometry include both
  quasinormal modes and total-transmission modes.  Sequences of these
  modes are parameterized by the angular momentum of the black hole.
  The quasinormal and total-transmission modes are usually distinct,
  having mode frequencies that are different at any given value of the
  angular momentum.  But a discrete and countably infinite subset of
  the left-total-transmission modes are simultaneously quasinormal
  modes.  Most of these special modes exist along previously unknown
  branches of the gravitational total-transmission modes.  In this
  paper, we give detailed plots of the total-transmission modes for
  harmonic indices $\ell=[2,7]$, with special emphasis given to the
  $m=0$ modes which all contain previously unknown branches.  All of
  these unknown branches have purely imaginary mode frequencies.  We
  find that as we approach the Schwarzschild limit along these new
  branches, the mode frequencies approach $-i\infty$ in stark contrast
  to the finite mode frequency obtained in the Schwarzschild limit
  along the previously known branches.  We explain when and why, at
  certain frequencies, the left-total-transmission modes are
  simultaneously quasinormal modes.  At these same frequencies, the
  right-total-transmission modes are missing.  We also derive analytic
  expressions for the asymptotic behavior of the total-transmission
  mode frequencies, and for the values of the angular momentum at
  which the modes are simultaneously quasinormal modes.
\end{abstract}


\maketitle

\section{Introduction}
\label{sec:introduction}

The Kerr geometry{\cite{kerr-1963}} is one of the most astrophysically
important solutions of Einstein's equations.  It represents an
isolated black hole with angular momentum, and the linear
perturbations of the Kerr geometry are used to understand the
late-time behavior of astrophysical events, such as the merger of
compact objects, that results in a black hole.  In particular, the
quasinormal modes (QNMs) of the Kerr spacetime are used to model the
ring-down portion of a gravitational waveform, for example from a
binary-black-hole
merger\cite{GW150914-2016,GW151226-2016,GW170104-2017,GW170814-2017}.

The gravitational modes of the Kerr geometry include not only the
QNMs, but also the total transmission modes (TTMs).  The QNMs are
defined by their allowing no disturbances to enter the system from
either infinity or from the black-hole boundary.  There are two types
of TTMs, each changing one of the conditions used in defining QNMs.
Left TTMs (TTM${}_{\rm L}$s) switch the boundary condition at the
black-hole boundary.  Essentially, TTM${}_{\rm L}$s represent
disturbances that can move from the vicinity of the black hole
toward spatial infinity without reflections that would propagate into
the black hole.  Right TTMs (TTM${}_{\rm R}$s) switch the boundary
condition at spatial infinity.  Essentially, TTM${}_{\rm R}$s
represent disturbances that can move from spatial infinity toward
the black hole without reflections that would propagate back toward
spatial infinity.  Switching both boundary conditions would represent
a bound state which does not exist for perturbations of the Kerr
geometry.  In this paper, we will be primarily interested in the
gravitational TTMs of the Kerr geometry.

The TTMs were first explored, in the context of algebraically special
modes, by Wald\cite{wald-1973} and in more detail by
Chandrasekhar\cite{chandra-1984}.  In the Schwarzschild limit, the
TTMs and their complex frequencies $\omega$ can be determined
analytically.  In general, the mode frequencies, $\omega$, are
functions of the angular momentum of the black hole, and must be
determined numerically.  The modes and their frequencies are
conveniently parameterized by the dimensionless angular momentum
$\bar{a}=a/M$, where $M$ is the mass of the black hole, and $J=aM$ is
the angular momentum.  In the Schwarzschild limit, $\bar{a}=0$, the TTM
frequencies $\omega(\bar{a})$ become
\begin{equation}\label{eq:alg-spec-sch}
  M\omega(0)\equiv\bar\Omega_\ell \equiv
  -\frac{i}{12}(\ell-1)\ell(\ell+1)(\ell+2),
\end{equation}
which are purely imaginary.  In general the mode frequencies
$\omega(\bar{a})$ are complex and the TTM${}_{\rm L}$s and TTM${}_{\rm
  R}$s share the same set of mode frequencies, although the modes
themselves differ.

The first numerical results for the mode frequencies of the TTMs were
included in Chandrasekhar's derivation of these algebraically special
perturbations\cite{chandra-1984}.  There, a few values were given for
each of the five modes for $\ell=2$.  More detailed plots for the
$\ell=2$ modes were shown by Onozawa\cite{onozawa-1997}.  More
recently, one of us reported the mode frequencies for both $\ell=2$
and $\ell=3$\cite{cook-zalutskiy-2014}.  Interestingly, numerical
investigations\cite{leaver-1985,onozawa-1997} showed that in the
Schwarzschild limit, certain QNM frequencies also approached the same
set of purely imaginary mode frequencies $\Omega_\ell$.  There was
considerable confusion about the nature of the modes with these
frequencies, but this confusion was set to rest by Massen van den
Brink\cite{van_den_brink-2000}.  He showed that the TTM${}_{\rm L}$
associated with a given $\bar\Omega_\ell$ was actually, simultaneously
a QNM.  At the same time, the corresponding solution thought to be a
TTM${}_{\rm R}$ was neither a TTM nor a QNM.  These conclusions are based
on a careful examination of the behavior of the modes at the
black-hole horizon (see Refs.~\cite{van_den_brink-2000} or
\cite{cook-zalutskiy-2016b} for more detail.)

Recent high-precision explorations of the gravitational QNMs of the
Kerr geometry found a large number of additional cases where sequences
of QNM frequencies $\omega(\bar{a})$ approached the negative
imaginary axis (NIA) for values of
$\bar{a}\ne0$\cite{cook-zalutskiy-2016b}.  The nature of the modes
associated with these new, purely imaginary mode frequencies was
explored\cite{cook-zalutskiy-2016a,cook-zalutskiy-2016b} using the
theory of Heun polynomials\cite{Heun-eqn}, expanding on the previous
work by Massen van den Brink\cite{van_den_brink-2000}.

In certain cases, the sequence of QNMs could have frequencies approach
the NIA, but could not exist with $\omega(\bar{a})$ precisely on the
NIA.  In other cases, all with $m=0$, the sequence of QNMs could
extend to have $\omega(\bar{a})$ on the NIA.  In this case, two
possible behaviors were seen.  Either the mode was simply a QNM, or it
was simultaneously a QNM and a TTM${}_{\rm L}$.  This latter case is
an extension of the behavior seen in the Schwarzschild limit (ie. the
modes with frequencies $\bar\Omega_\ell$.)

Let us reconsider this with emphasis on the behavior of the TTMs.  For
generic values of $\bar{a}$, the TTM${}_{\rm L}$s and TTM${}_{\rm R}$s
share the same set of mode frequencies.  In the Schwarzschild limit,
$\bar{a}=0$, the various $m$ modes are degenerate.  Also, the
TTM${}_{\rm L}$s are simultaneously QNMs while the solutions that
should be TTM${}_{\rm R}$s are missing.  It was seen in
Ref.~\cite{cook-zalutskiy-2016b} that for certain values of
$\bar{a}>0$ for $\ell=2\text{---}4$ and $m=0$, certain QNMs on the NIA
were also simultaneously TTM${}_{\rm L}$s.  At these same values of
$\bar{a}$, the corresponding TTM${}_{\rm R}$s did not exist.  This
behavior occurred at certain points along $m=0$ TTM sequences within
the range $0\le\bar{a}<\bar{a}_{\mbox{\tiny crit}}$ (where
$\bar{a}_{\mbox{\tiny crit}}$ is a function of $\ell$), and within
this range $\omega(\bar{a})$ is purely imaginary.  For
$\bar{a}_{\mbox{\tiny crit}}<\bar{a}\le1$, the $m=0$ TTM sequences of
mode frequencies move off of the NIA, obtaining general complex
values.

However, we found\cite{cook-zalutskiy-2016b} that for each
$\ell=2\text{---}4$ and $m=0$, there are a seemingly countably
infinite number of QNMs with $\bar{a}<\bar{a}_{\mbox{\tiny crit}}$
that are also simultaneously TTM${}_{\rm L}$s (and at which the
TTM${}_{\rm R}$s do not exist), but which do not correspond to know
segments of the TTM sequences.  This indicated that a previously
unknown branch exists for each of the $m=0$ TTM sequences.  In this
paper, we verify this conjecture by explicitly computing these
sequences for $\ell=2\text{---}7$.

Interestingly, the mode frequencies of this new branch of the $m=0$
TTMs approach $-i\infty$ as $\bar{a}\to0$.  We find that it is
possible to construct an analytic expression for the asymptotic
behavior of these branches.  Doing so required that we find an analytic
expression for the separation constant $\scA[]{s}{\ell m}{c}$ of the
spin-weighted spheroidal harmonics in the asymptotic prolate case
(large purely imaginary $c$) for $m=0$.  To our knowledge this has
been known only to leading order for $s=\pm2$.

This paper proceeds as follows.  In Sec.~\ref{sec:methods}, we will
briefly review the methods used to construct TTMs of the Kerr
geometry.  In Sec.~\ref{sec:numerical_results}, we will examine
numerical results for the TTMs.  In
Sec.~\ref{sec:asymptotic_behavior}, we will find an asymptotic
expansion for both the mode frequencies, $\omega(\bar{a})$ of the TTMs
and for the separation constant $\scA[]{s}{\ell m}{c}$ of the
spin-weighted spheroidal harmonics.  Finally, in
Sec.~\ref{sec:summary}, we will review the prior results from
Ref.\cite{cook-zalutskiy-2016b} that describe where along the TTM
sequences the modes change their behavior, and end with some final
discussion.

\section{Methods}
\label{sec:methods}

Modes of the Kerr geometry can be obtained by solving the Teukolsky
master equation with appropriate boundary conditions.  Our approach
for obtaining the QNMs of the Kerr geometry is outlined in detail in
Refs.~\cite{cook-zalutskiy-2014} and \cite{cook-zalutskiy-2016b}.
Our methods for obtaining the TTMs are essentially the same, but with
a few necessary differences.

In vacuum, the Teukolsky master equation separates using
\begin{equation}\label{eq:Teukolsky_separation_form}
  {}_s\psi(t,r,\theta,\phi) = e^{-i\omega{t}} e^{im\phi}S(\theta)R(r).
\end{equation}
The radial function $R(r)$ then satisfies the radial Teukolsky
equation
\begin{subequations}
\begin{align}\label{eqn:radialR:Diff_Eqn}
\Delta^{-s}\frac{d}{dr}&\left[\Delta^{s+1}\frac{dR(r)}{dr}\right]
 \\
&+ \left[\frac{K^2 -2is(r-M)K}{\Delta} + 4is\omega{r} - \lambdabar\right]R(r)=0,
\nonumber
\end{align}
where
\begin{align}
  \Delta &\equiv r^2-2Mr+a^2, \\
  K &\equiv (r^2+a^2)\omega - am, \\
  \lambdabar &\equiv \scA{s}{\ell{m}}{a\omega} + a^2\omega^2 - 2am\omega,
\end{align}
\end{subequations}
and Boyer-Lindquist coordinates are used.  $\scA{s}{\ell{m}}{a\omega}$
is the angular separation constant associated with the angular
Teukolsky equation governing $S(\theta)$.  With $x=\cos\theta$, the
function $S(\theta)=\swS{s}{\ell{m}}{x}{a\omega}$ is the spin-weighted
spheroidal function satisfying
\begin{align}\label{eqn:swSF_DiffEqn}
\partial_x \Big[ (1-x^2)\partial_x [\swS{s}{\ell{m}}{x}{c}]\Big] & \nonumber \\ 
    + \bigg[(cx)^2 - 2 csx + s& + \scA{s}{\ell m}{c}   \\ 
      & - \frac{(m+sx)^2}{1-x^2}\bigg]\swS{s}{\ell{m}}{x}{c} = 0,
\nonumber
\end{align}
where $c\ (=a\omega)$ is the oblateness parameter and $m$ the
azimuthal separation constant.  Finally, $\ell$ is the harmonic mode
index which labels the elements in the set of eigensolutions of the
angular equation for fixed values of $s$, $m$, and $c$.

Both Eqs.~(\ref{eqn:radialR:Diff_Eqn}) and (\ref{eqn:swSF_DiffEqn})
are examples of the general class of confluent Heun equations.  We
have found it particularly useful to work with the radial equation,
Eq.~(\ref{eqn:radialR:Diff_Eqn}), within the context of confluent Heun
theory\cite{Heun-eqn}.  Following Borissov and
Fiziev\cite{Fiziev-2009b}, the radial equation can be written in the
nonsymmetrical canonical form of the confluent Heun equation in 8
different ways, depending on the choice of three parameters.  Following
Ref.~\cite{cook-zalutskiy-2014}, we denote these three parameters as
$\bar\zeta$, $\xi$, and $\eta$.  Each parameter has two possible
choices:
\begin{equation}\label{eq:Teukolsky_Heun_parameters}
  \bar\zeta=\bar\zeta_\pm, \qquad
  \xi=\xi_\pm, \qquad\mbox{and}\qquad
  \eta=\eta_\pm.
\end{equation}
These three parameters are associated with the regular behavior of
solutions of the confluent Huen equation at the three singular points.
See Ref.~\cite{cook-zalutskiy-2014} for details.  Regular singular
points occur at the Cauchy and event horizons, with the choice of
$\xi$ fixing the behavior at the event horizon, and $\eta$ fixing the
behavior at the Cauchy horizon.  The singular point at infinity is
irregular, and is fixed by the choice of $\bar\zeta$.  See Sec.~II.C
of Ref. ~\cite{cook-zalutskiy-2014} for details.

There are two local Frobenius solutions at each singular point, but
more interesting are the solutions known as {\em confluent Heun
  functions} and {\em confluent Heun polynomials}.  Confluent Heun
functions are solutions that are simultaneously Frobenius solutions at
two adjacent singular points.  For our purposes, they arise when we
fix boundary conditions at two singular points and find values of some
parameter that allows both boundary conditions to be satisfied.  In
general, confluent Heun functions are infinite-series solutions.
Confluent Heun polynomials are an important subset of the confluent
Heun functions where the solutions are simultaneously Frobenius
solutions of all three singular points.  In this case, the
infinite-series solutions truncate to yield a polynomial solution.

By making an additional coordinate transformation and an appropriate
redefinition of the solution function (see Sec.~II.C.1 of
Ref. ~\cite{cook-zalutskiy-2014} for details), the resulting equation
takes a form where the only solution that is finite at both singular
points is the solution that satisfies the desired boundary conditions.
In this form, the desired confluent Heun functions can be found using
Leaver's continued-fraction approach\cite{leaver-1985}.  As outlined
in Sec.~IV.B of Ref.\cite{cook-zalutskiy-2016b}, the choices
$\bar\zeta=\bar\zeta_+$ and $\xi=\xi_\minus$ (see Eqns.~(19a) and
(19b) in \cite{cook-zalutskiy-2016b}) together with a specific
redefinition of the solution function allows the continued-fraction
method to find confluent Heun functions which are QNMs.  The entire
procedure is extensible to finding TTMs by simply changing the choice
of $\xi$ to $\xi_+$, in which case the confluent Heun functions are
TTM${}_{\rm L}$s.  To find TTM${}_{\rm R}$s, we instead change the
choice of $\bar\zeta$ to $\bar\zeta_-$.

In more detail, the continued-fraction solution of Eq.~(30) in
Ref.~\cite{cook-zalutskiy-2014} is unchanged except for how the
choices for the $\xi$ and $\bar\zeta$ parameters affect the evaluation
of the coefficients in Eqs.~(31a--e) and (38a--c) of
\cite{cook-zalutskiy-2014}.  The only changes in the discussion of the
existence of minimal solutions of the continued fraction found in
Ref.~\cite{cook-zalutskiy-2016b} come from the fact that for
TTM${}_{\rm R}$s, the branch cut of $u_1(\bar\omega)$ (see discussion
below Eq.~(25b)) is along the positive imaginary axis, which means
that ${\rm Re}(u_1(\bar\omega))\ne0$ on the NIA.  This means that
the continued fraction can be used to determine the TTM${}_{\rm R}$
frequencies on the NIA, in contrast with the QNM and TTM${}_{\rm L}$
cases.

In addition to the confluent Heun function mode solutions of the
Teukolsky radial equation, there also exist confluent Heun polynomial
mode solutions.  For TTMs, the derivation of the polynomial modes can
be found in Sec.~III.B of Ref.\cite{cook-zalutskiy-2014}.  There, we
find that the condition for the existence of confluent Heun polynomial
solutions is equivalent to the vanishing of the magnitude squared of
the Starobinsky constant\cite{wald-1973,chandra-1984}.  We write the
magnitude squared of the Starobinsky constant as
\begin{align}\label{eq:Starobinsky_const}
  |\mathcal{Q}|^2 =\lambdabar^2(\lambdabar+2)^2 
  &+ 8\lambdabar\bar{a}\bar\omega\left(
      6(\bar{a}\bar\omega+m) -5\lambdabar(\bar{a}\bar\omega-m)\right)
      \nonumber\\ \mbox{}&
      + 144\bar\omega^2\left(1+\bar{a}^2(\bar{a}\bar\omega-m)^2\right),
\end{align}
where $\bar\omega\equiv M\omega$ is the dimensionless mode frequency.
For TTM${}_{\rm L}$s, take $s=-2$ and
\begin{equation}
  \lambdabar=\lambdabar_\minus\equiv \scA{-2}{\ell{m}}{\bar{a}\bar\omega}
     + \bar{a}^2\bar\omega^2 - 2m\bar{a}\bar\omega.
\end{equation}
For TTM${}_{\rm R}$s, take $s=+2$ and
\begin{equation}
  \lambdabar=\lambdabar_+\equiv \scA{2}{\ell{m}}{\bar{a}\bar\omega}
     + \bar{a}^2\bar\omega^2 - 2m\bar{a}\bar\omega+4.
\end{equation}
However, because
$\scA{-s}{\ell{m}}{\bar{a}\bar\omega}=\scA{s}{\ell{m}}{\bar{a}\bar\omega}+2s$,
it follows that $\lambdabar_+=\lambdabar_\minus$ and we find that the
TTM${}_{\rm L}$ and TTM${}_{\rm R}$ algebraically special modes will
share the same frequency spectrum.  For $\bar{a}>0$, numerical values
for the TTM mode frequencies $\bar\omega$ are obtained by finding
roots of Eq.~(\ref{eq:Starobinsky_const}) where values for
$\scA{s}{\ell{m}}{\bar{a}\bar\omega}$ are obtained by solving the
angular Teukolsky equation, Eq.~(\ref{eqn:swSF_DiffEqn}), using the
spectral solver described in Sec.~II.D.1 of
Ref.\cite{cook-zalutskiy-2014}.

For QNMs, the derivation of the polynomial modes can be found in
Sec.~IV.C of Ref.~\cite{cook-zalutskiy-2016b}.  Confluent Heun
polynomial QNM solutions can exist in two classes distinguished by the
choice of the $\eta$ parameter.  For both classes, there is a
necessary, but not sufficient condition placed on the value of
$\bar\omega$ in order for a polynomial QNM solution to exist.  A
corresponding constraint on the TTM frequencies does not exist.  The
constraints on the values of $\bar\omega$ for polynomial QNM solutions
can be written as
\begin{equation}\label{eq:omega_plus}
\bar\omega=\bar\omega_+ \equiv
\frac{\bar{a}m-iN_+\sqrt{1-\bar{a}^2}}{2(1+\sqrt{1-\bar{a}^2})},
\end{equation}
or
\begin{equation}\label{eq:omega_minus}
\bar\omega=\bar\omega_\minus \equiv
-i\frac{N_\minus}4,
\end{equation}
where $N_+\ge s+1$ and $N_\minus\ge1$ take on integer values.  The
necessary and sufficient condition for a polynomial QNM solution to
exist is referred to as the $\Delta_{q+1}=0$ condition, and is
described in detail in Sec.~IV.C.1 of
Ref.~\cite{cook-zalutskiy-2016b}.  The vanishing of the magnitude
squared of the Starobinsky constant is a special case of the
$\Delta_{q+1}=0$ condition.

As discussed in Sec.~IV.D of Ref.~\cite{cook-zalutskiy-2016b}, for
potential QNM polynomial modes, we must pay particular attention to
the behavior of the solution at the event horizon.  The constraint
that $\bar\omega=\bar\omega_+$ is precisely the condition that the
roots of the indicial equation differ by an integer for local
Frobenius solutions at the event horizon.  Summarizing the results of
Ref.~\cite{cook-zalutskiy-2016b}, all of the potential polynomial QNM
modes found have $m=0$.  The majority of potential QNM modes
satisfying $\bar\omega=\bar\omega_+$ are {\em miraculous}.  That is,
such a mode is neither QNM nor TTM${}_{\rm L}$\footnote{It cannot be
  TTM${}_{\rm R}$ because of the boundary condition at infinity.}.
However, a subset of potential QNM modes are {\em anomalous}.  That
is, such a mode is simultaneously a QNM and a TTM${}_{\rm L}$.  In
fact, these anomalous QNM solutions also satisfy the condition
$|\mathcal{Q}|^2=0$ for polynomial TTM${}_{\rm L}$s.  Finally, while
not of direct interest to the remainder of this paper, all of the
potential QNM modes satisfying $\bar\omega=\bar\omega_\minus$ have
proven to be generic.  That is, the roots of the indicial equation do
not differ by an integer, and the solutions are simply QNMs.

\section{Numerical results.}
\label{sec:numerical_results}

\begin{figure}
\includegraphics[width=\linewidth,clip]{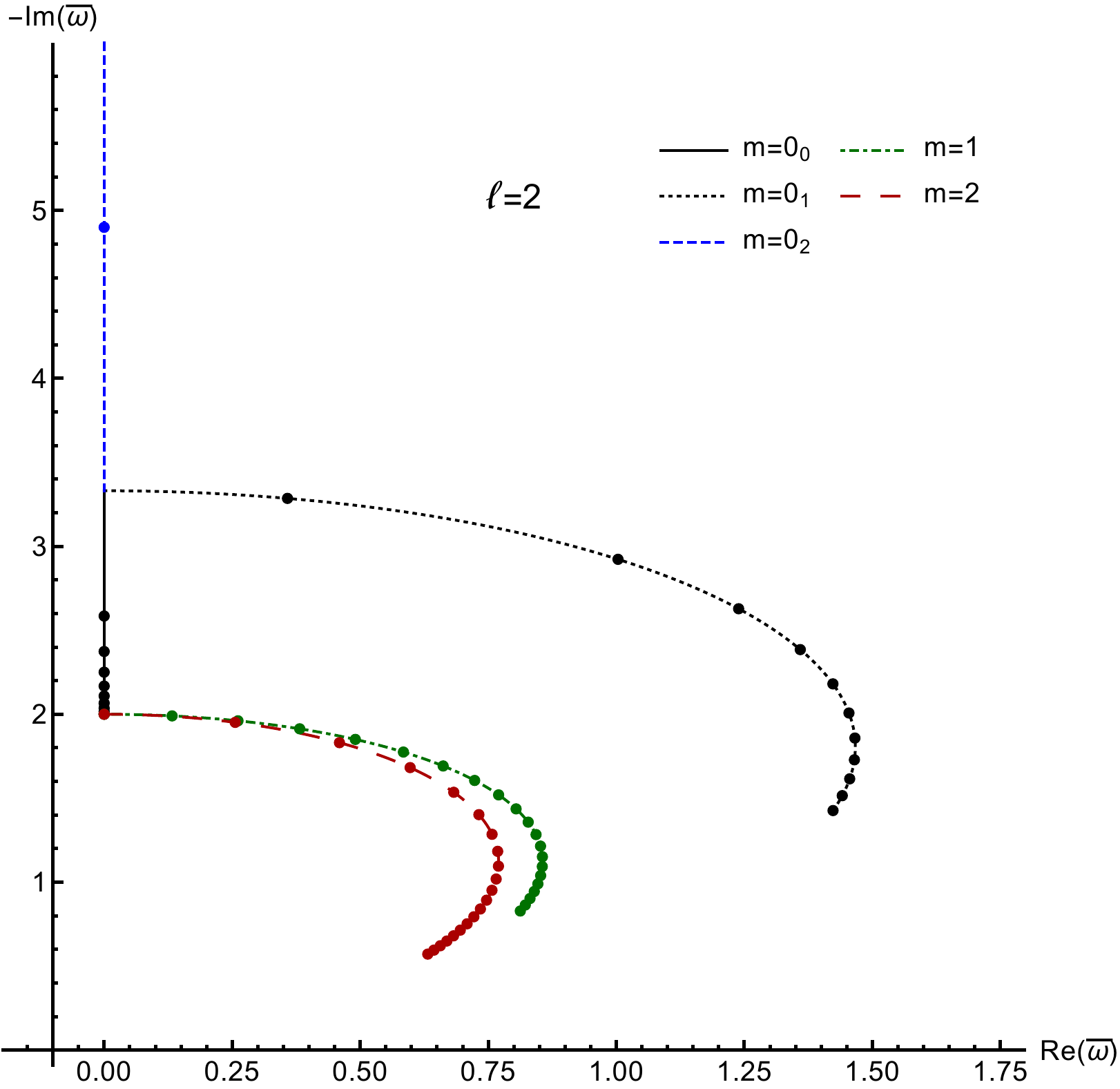}
\caption{\label{fig:TTMl2} The complex frequency $\bar\omega$ is
  plotted for the Kerr TTM $\ell=2$ modes.  Note that the imaginary
  axis is inverted.  Each sequence covers the range $0\le\bar{a}\le1$,
  with markers on each sequence denoting a change in $\bar{a}$ of
  $0.05$.  The sequences labeled $m=0_0$ and $0_1$ are the previously
  known segments of the $m=0$ sequence.  The segment labeled $m=0_2$
  is the previously unknown branch of the $m=0$ sequence.}
\end{figure}

The first numerical results for the mode frequencies of the TTMs were
included in Chandrasekhar's derivation of these algebraically special
perturbations\cite{chandra-1984}.  There, a few values were given for
each of the five modes for $\ell=2$.  More detailed plots for the
$\ell=2$ modes were shown by Onozawa\cite{onozawa-1997}.  More
recently, we reported the mode frequencies for both $\ell=2$ and
$\ell=3$\cite{cook-zalutskiy-2014}.  Figure~\ref{fig:TTMl2} shows the
various $\ell=2$ TTM mode frequency sequences.  All start at the
Schwarzschild limit at the point $\bar\omega=-2i$.  The $m=1$ and
$m=2$ sequences extent rightward from this point.  The $m=0$ sequence
first extends up the NIA.  This portion of the sequence is labeled
$m=0_0$ in the figure, and covers the range
$0\le\bar{a}\lesssim0.494445955$.  For
$0.494445955\lesssim\bar{a}\le1$, the $m=0$ sequence abruptly turns
off of the NIA and extends rightward.  This segment of the sequence is
labeled $m=0_1$ in the figure.  The segment of the sequence labeled
$m=0_2$ is the previously unknown segment of the sequence first
noticed in Ref.~\cite{cook-zalutskiy-2016b}.  It also covers the range
$0\le\bar{a}\lesssim0.494445955$ covered by the $m=0_0$ segment, but
$\lim_{\bar{a}\to0}\bar\omega=-i\infty$.

\begin{figure}[h]
\begin{tabular}{cc}
\includegraphics[width=0.95\linewidth,clip]{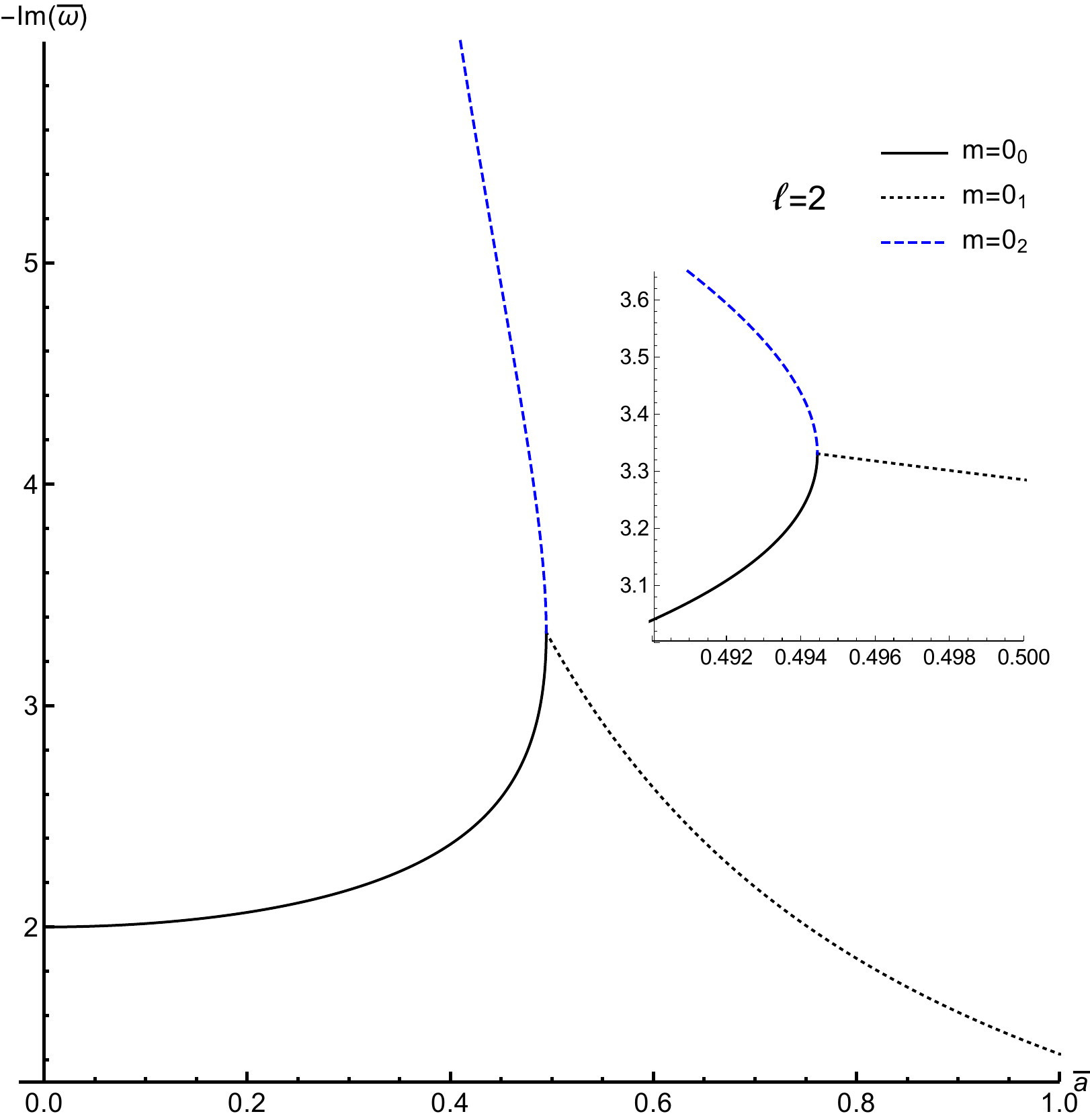} \\
\includegraphics[width=0.95\linewidth,clip]{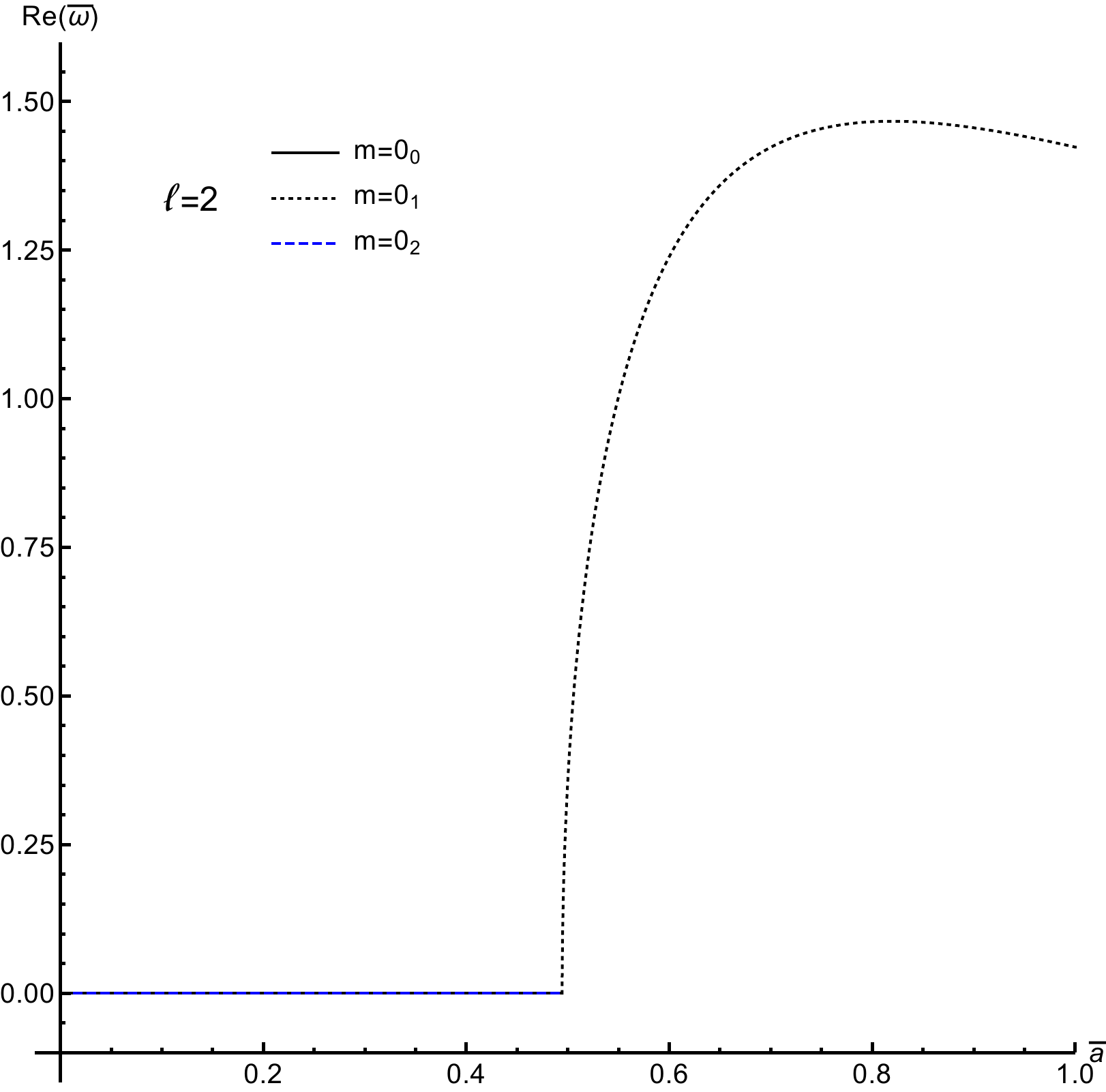}
\end{tabular}
\caption{\label{fig:TTMl2m0ImRe} The real and imaginary parts of the
  complex frequency $\bar\omega$ is plotted as a function of $\bar{a}$
  for the Kerr TTM $\ell=2$, $m=0$ sequence.  Note that the axis is
  inverted for the ${\rm Im}(\bar\omega)$.  For the ${\rm
    Im}(\bar\omega)$, the $m=0_2$ segment continues towards $-i\infty$
  and asymptotes to the NIA.  The inset in the first plot shows the
  behavior near the critical value of $\bar{a}$.  For the ${\rm
    Re}(\bar\omega)$, the $m=0_0$ and $0_2$ segments overlap with
  ${\rm Re}(\bar\omega)=$ along these segments.}
\end{figure}

The behavior of the $\ell=2$, $m=0$ sequence can be more fully
understood by plotting the real and imaginary parts of $\bar\omega$
separately as functions of $\bar{a}$.  Figure~\ref{fig:TTMl2m0ImRe}
shows the 3 different $m=0$ segments as functions of $\bar{a}$.  Note
that for the ${\rm Im}(\bar\omega)$, the $m=0_0$ segment continues
smoothly into the $m=0_2$ segment at the critical value of $\bar{a}$
(where $d\bar\omega/d\bar{a}=\infty$), while the $m=0_1$ segment
emerges discontinuously from this point.

\begin{figure}[h]
\begin{tabular}{cc}
\includegraphics[width=0.95\linewidth,clip]{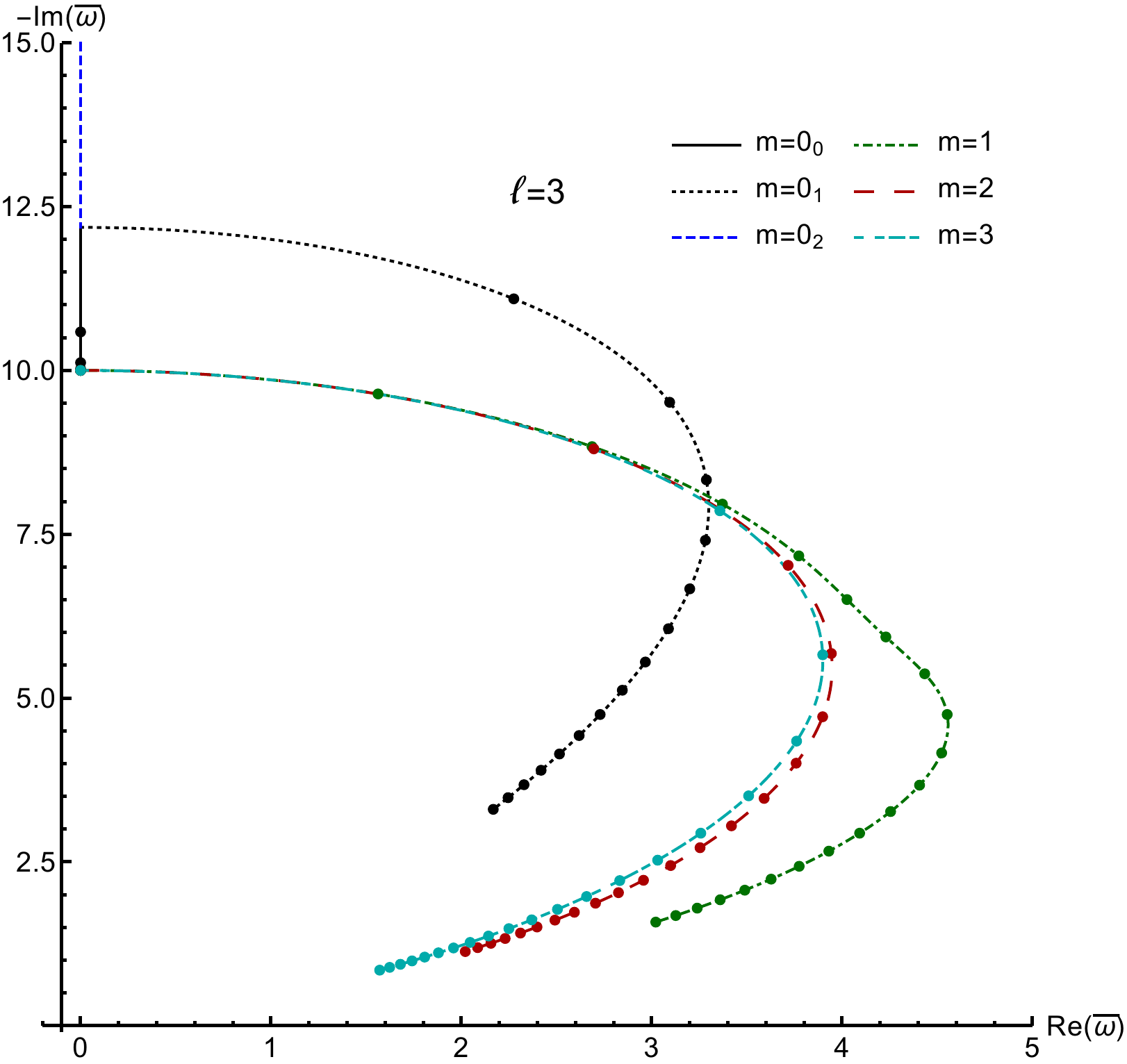} \\
\includegraphics[width=0.95\linewidth,clip]{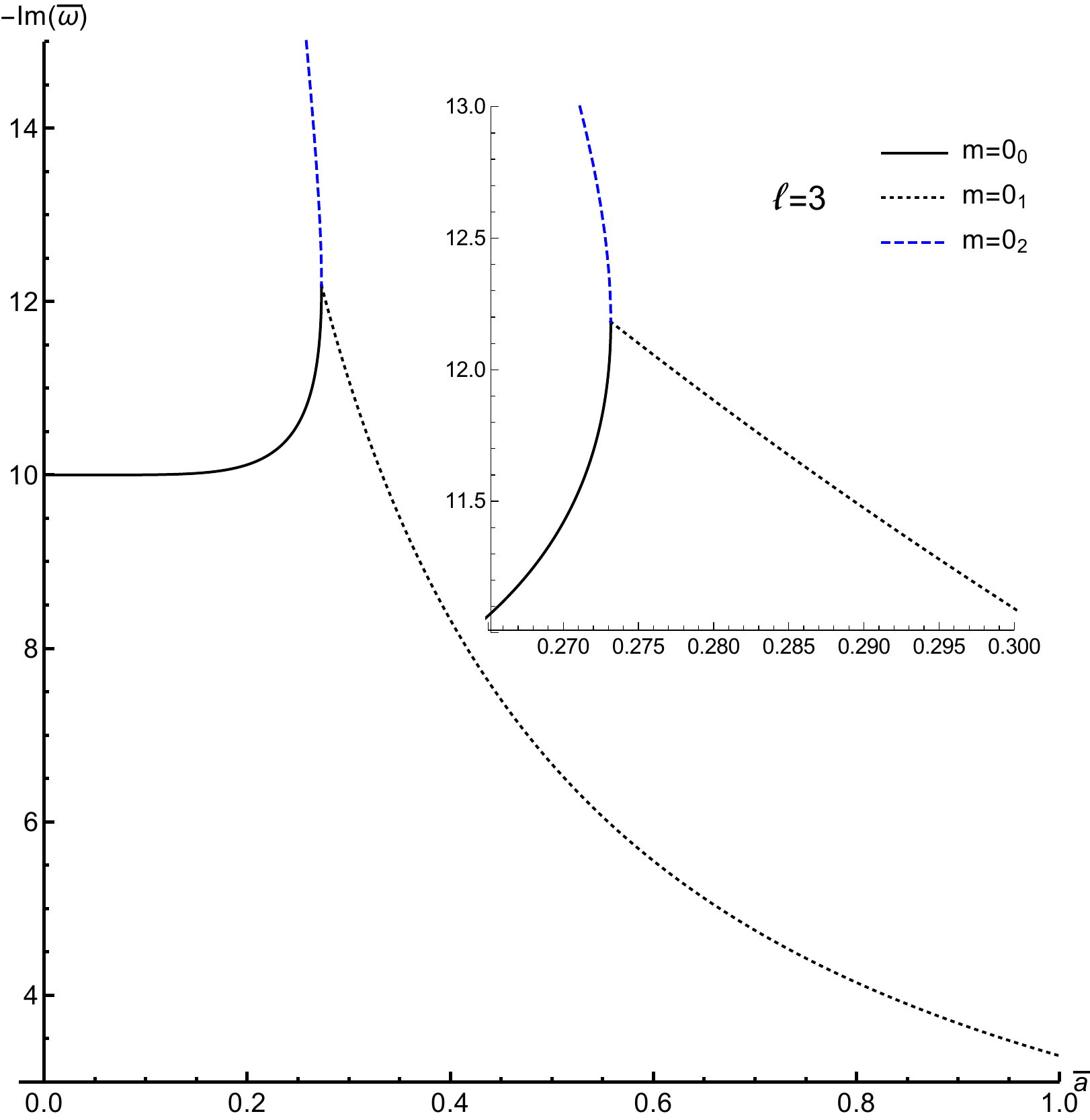}
\end{tabular}
\caption{\label{fig:TTMl3m} The first plot shows the complex frequency
  $\bar\omega$ for all of the TTM $\ell=3$ sequences.  The second plot
  shows the ${\rm Im}(\bar\omega)$ as a function of $\bar{a}$ for the
  Kerr TTM $\ell=3$, $m=0$ sequence.  Note that the ${\rm
    Im}(\bar\omega)$ axis is inverted for both plots.  The inset in
  the second plot shows the behavior near the critical value of
  $\bar{a}$.  The ${\rm Im}(\bar\omega)$ for the $m=0_2$ segment
  continues towards $-i\infty$ and asymptotes to the NIA.}
\end{figure}

\begin{figure}[h]
\begin{tabular}{cc}
\includegraphics[width=0.95\linewidth,clip]{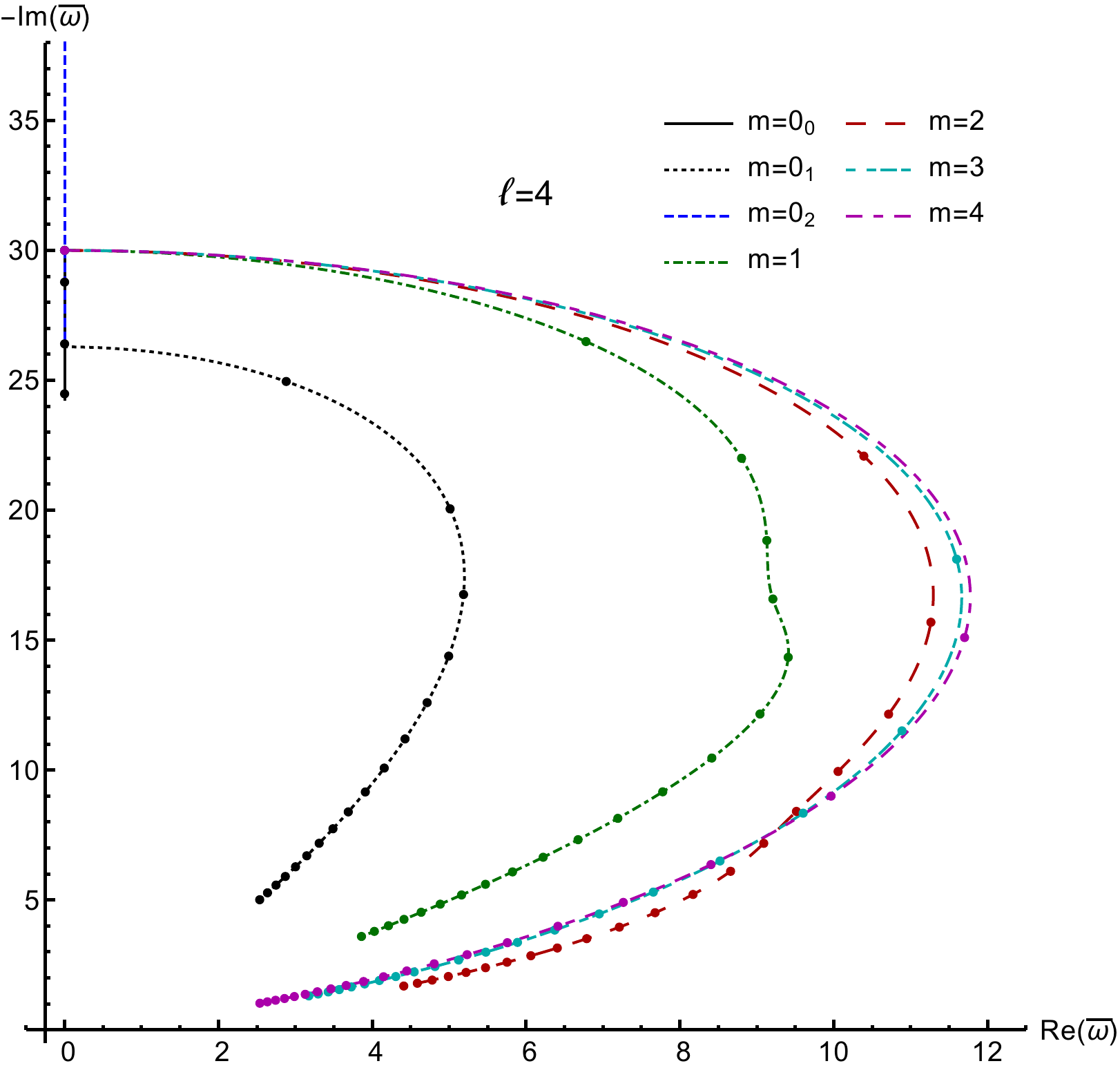} \\
\includegraphics[width=0.95\linewidth,clip]{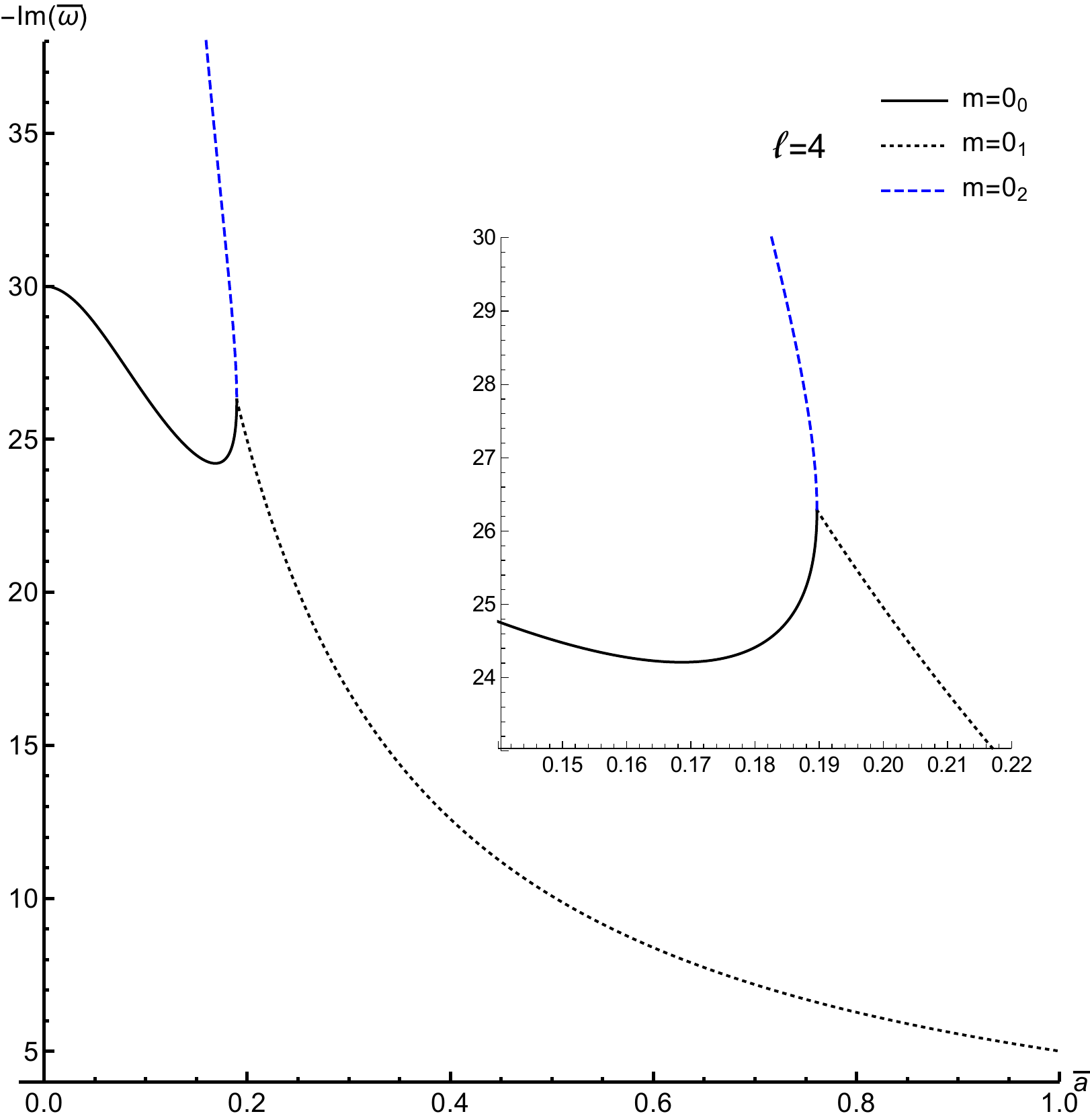}
\end{tabular}
\caption{\label{fig:TTMl4m} The TTMs for $\ell=4$.  See the caption to
  Fig.~\ref{fig:TTMl3m} for details.}
\end{figure}

\begin{figure}[h]
\begin{tabular}{cc}
\includegraphics[width=0.95\linewidth,clip]{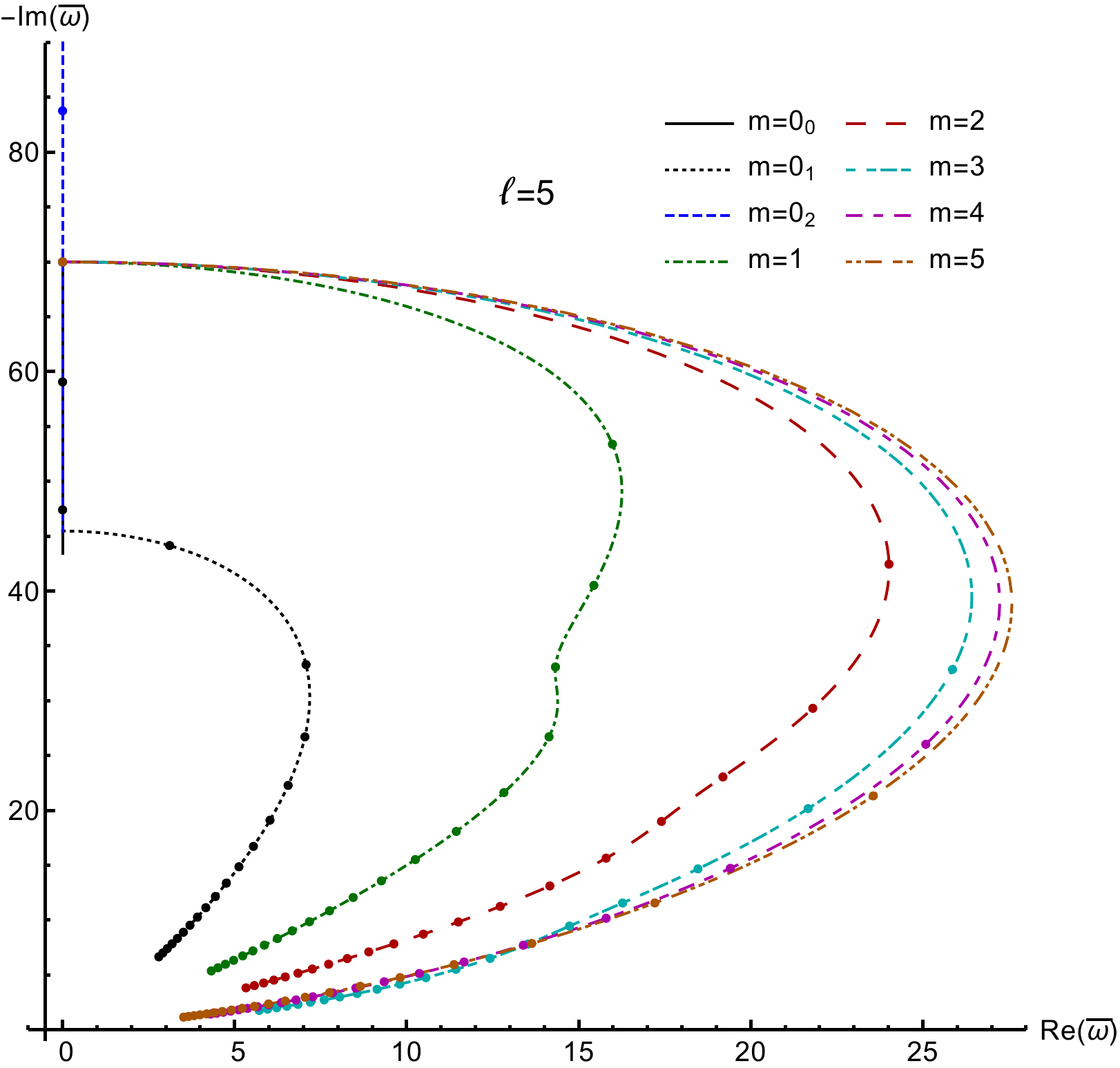} \\
\includegraphics[width=0.95\linewidth,clip]{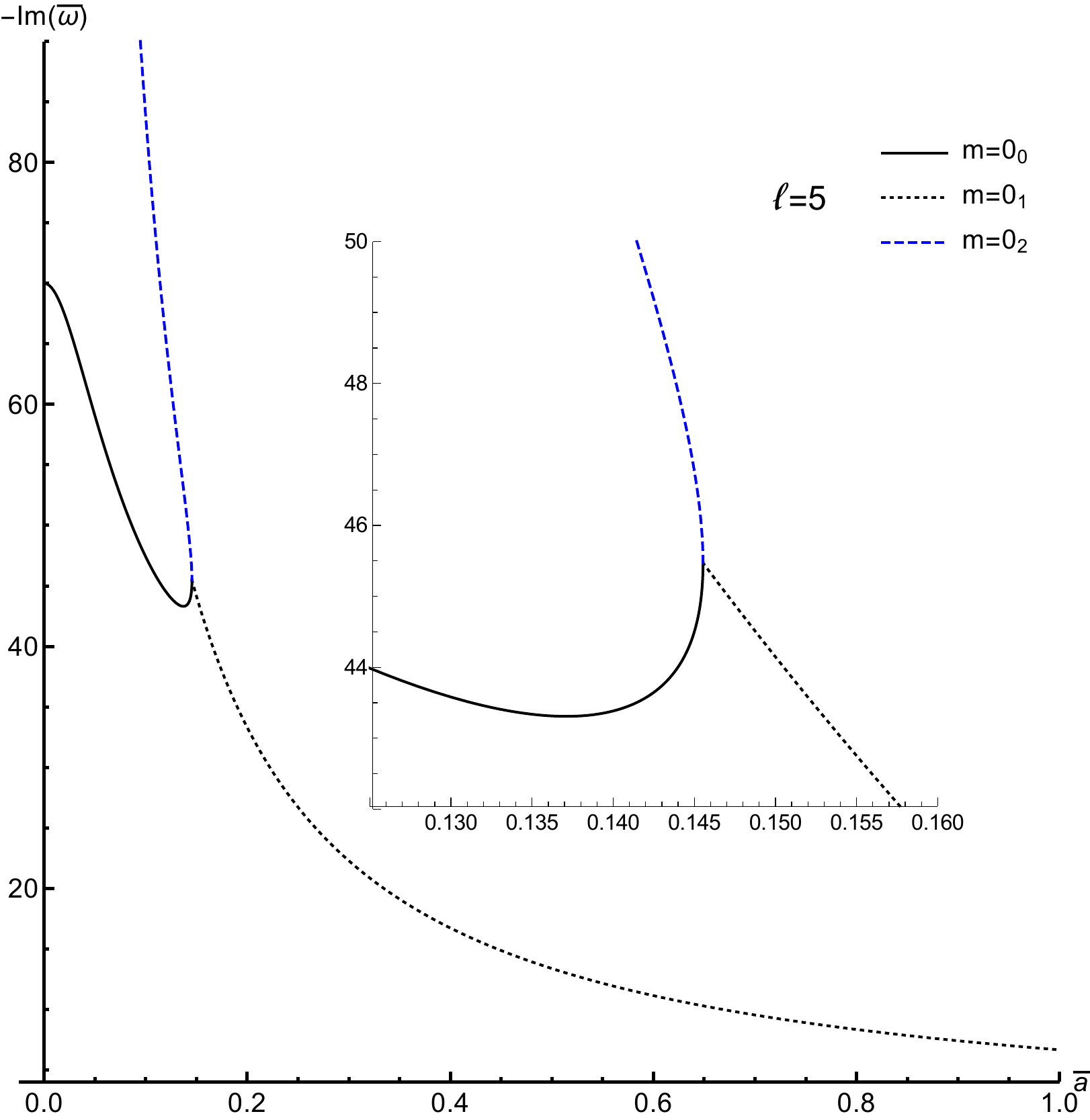}
\end{tabular}
\caption{\label{fig:TTMl5m} The TTMs for $\ell=5$.  See the caption to
  Fig.~\ref{fig:TTMl3m} for details.}
\end{figure}

\begin{figure}[h]
\begin{tabular}{cc}
\includegraphics[width=0.95\linewidth,clip]{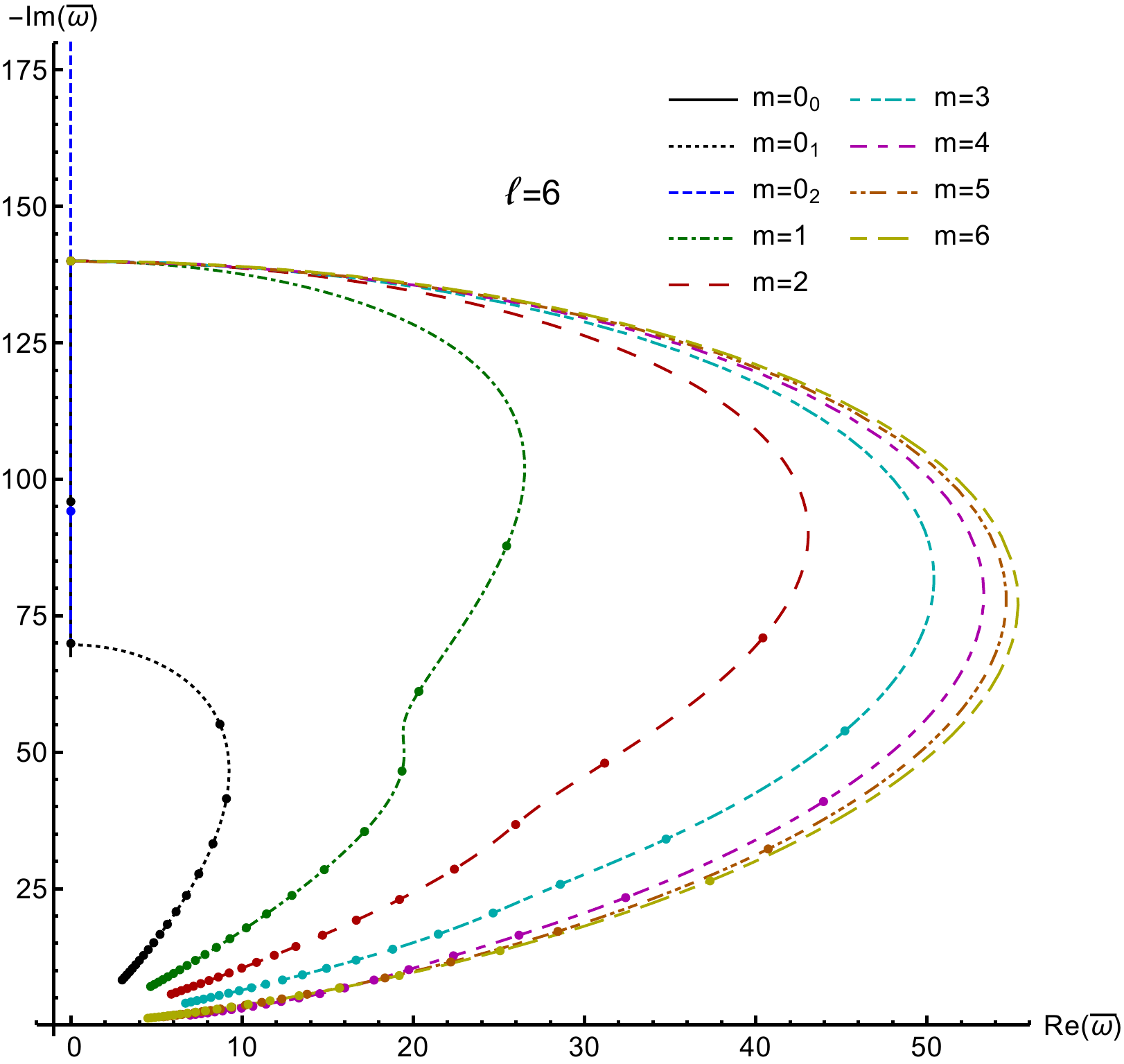} \\
\includegraphics[width=0.95\linewidth,clip]{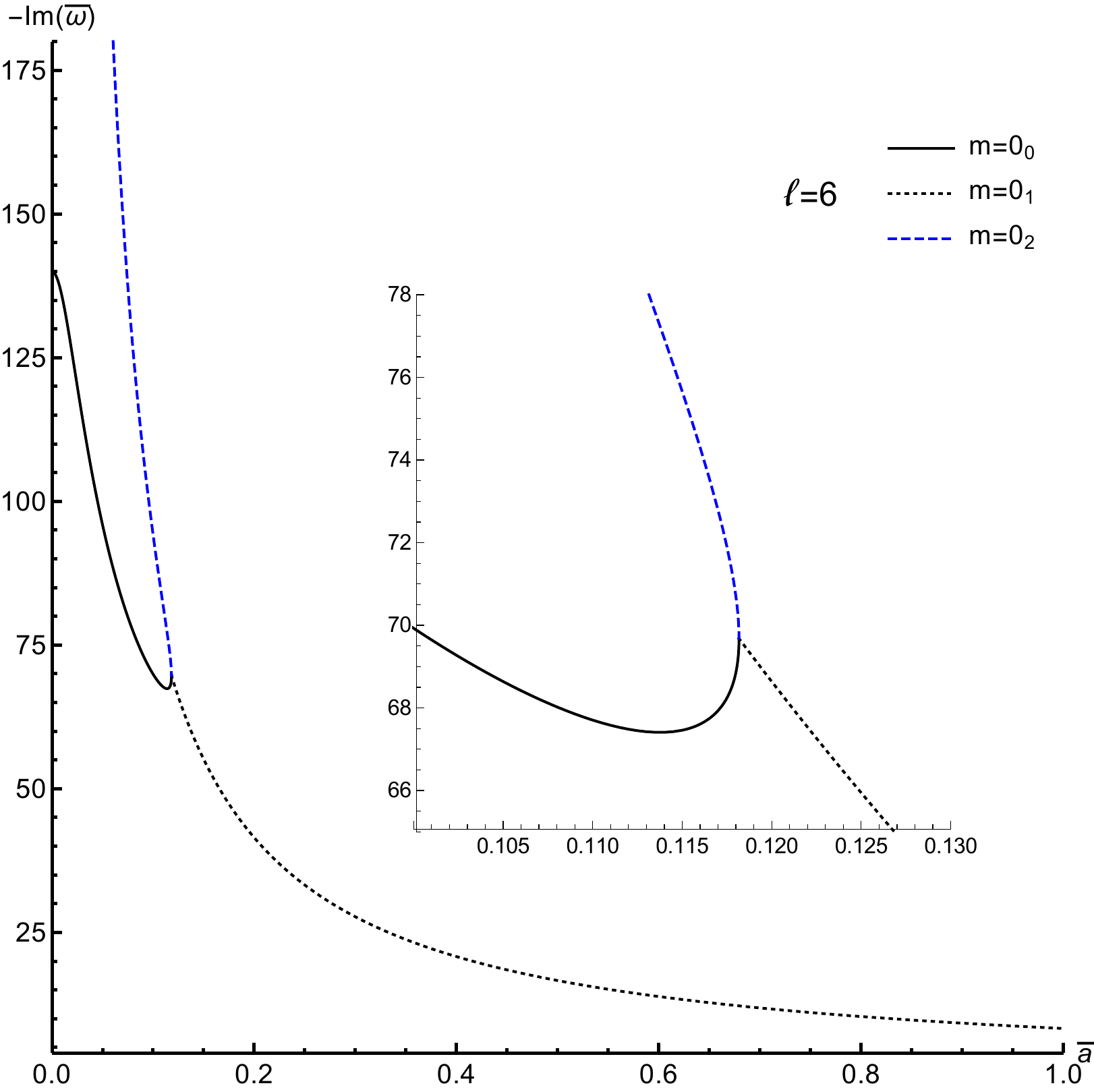}
\end{tabular}
\caption{\label{fig:TTMl6m} The TTMs for $\ell=6$.  See the caption to
  Fig.~\ref{fig:TTMl3m} for details.}
\end{figure}

\begin{figure}[h]
\begin{tabular}{cc}
\includegraphics[width=0.95\linewidth,clip]{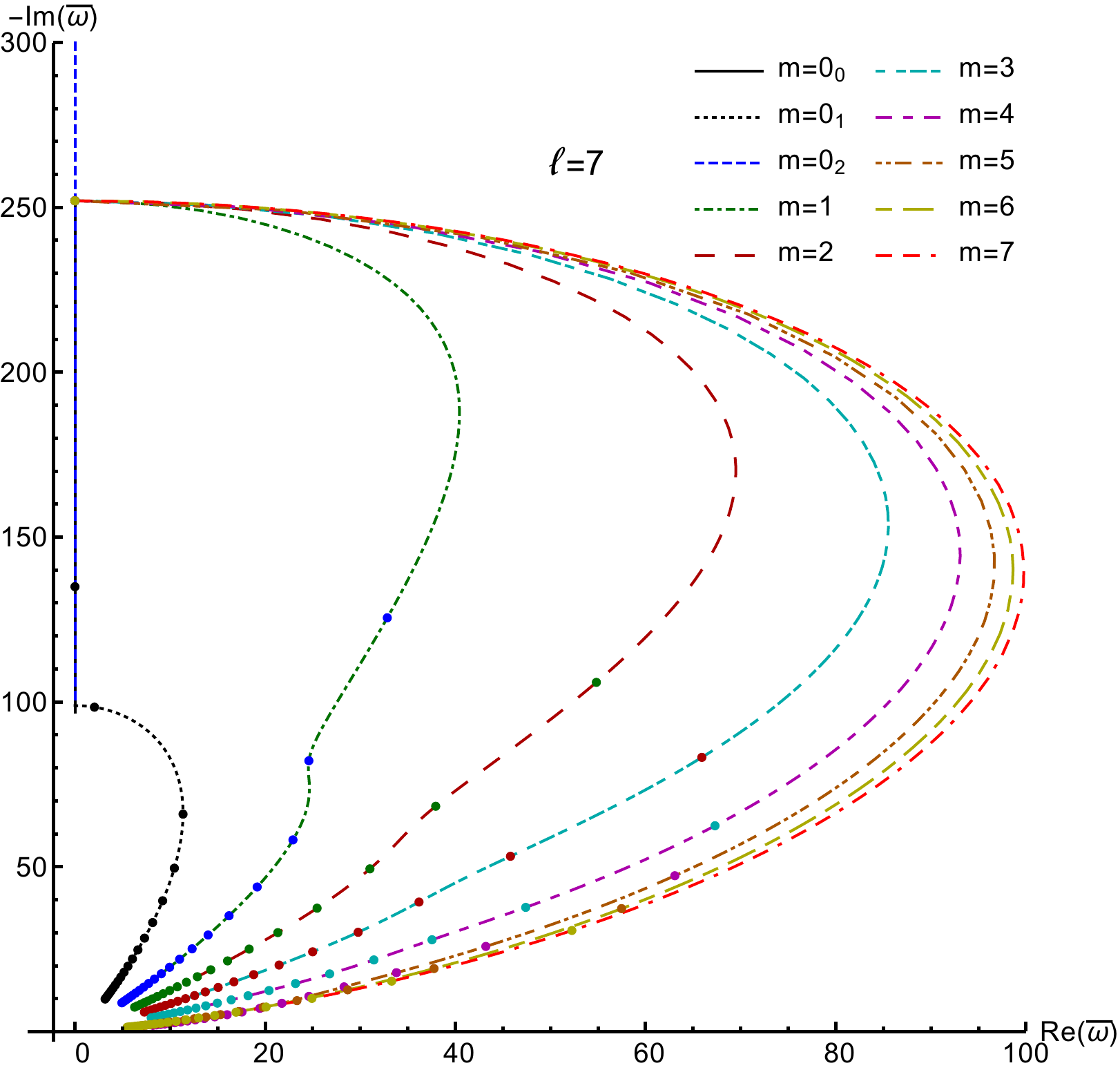} \\
\includegraphics[width=0.95\linewidth,clip]{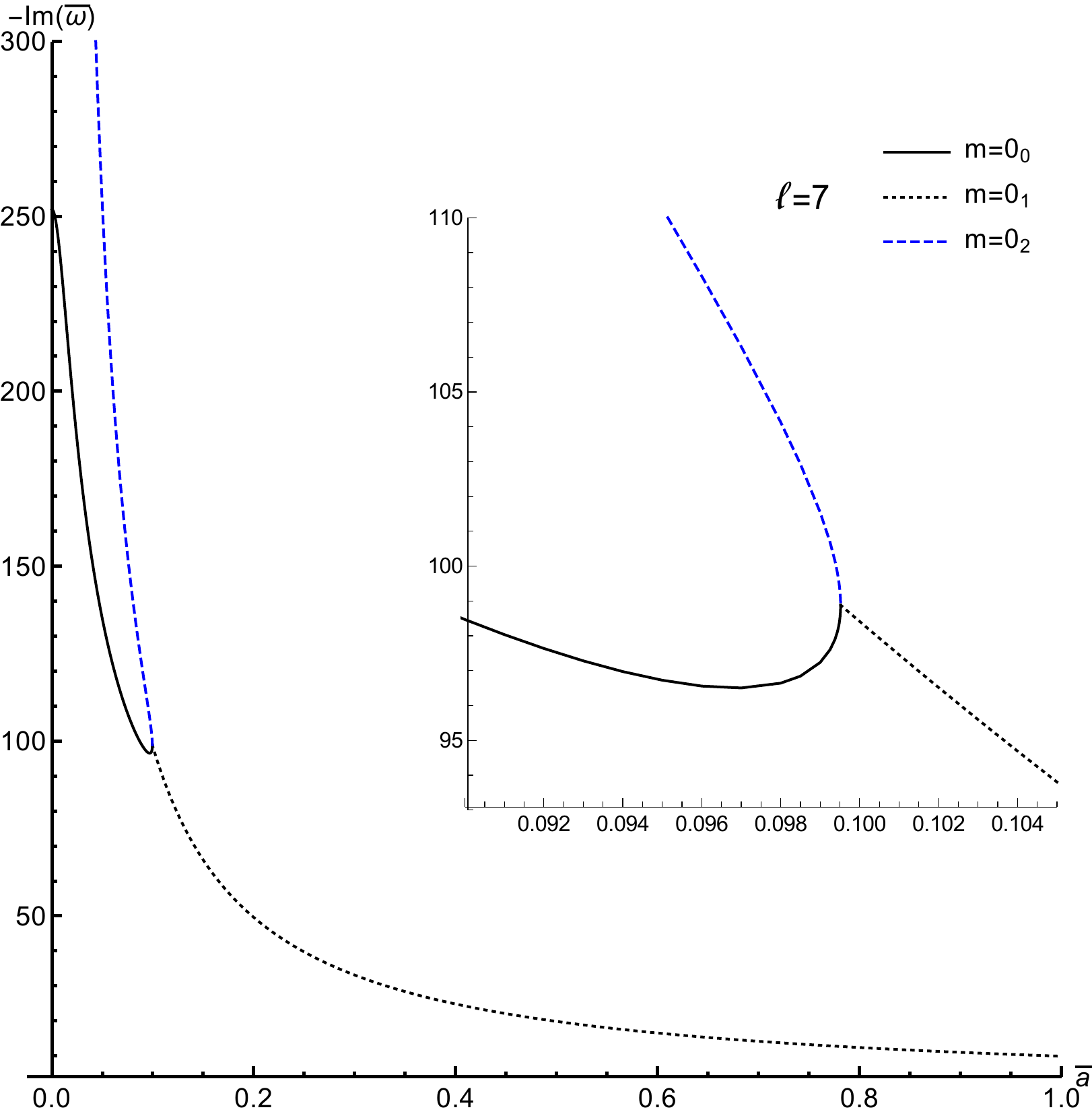}
\end{tabular}
\caption{\label{fig:TTMl7m} The TTMs for $\ell=7$.  See the caption to
  Fig.~\ref{fig:TTMl3m} for details.}
\end{figure}

The behavior for $\ell>2$ is similar.  Figures
\ref{fig:TTMl3m}--~\ref{fig:TTMl7m} show sequences of the complex
frequency $\bar\omega$ for all TTMs for $\ell=3$---$7$, and the ${\rm
  Im}(\bar\omega)$ as a function of $\bar{a}$ for each $m=0$ mode
sequence.

From fundamental symmetries of the radial and angular Teukolsky equations,
it is well known that mode frequencies come in pairs related by
\begin{equation}
  \label{eqn:freqsymmetry}
  \omega_{\ell -m}(a) = -\omega^*_{\ell m}(a).
\end{equation}
For QNMs, there are two complete sets of modes.  That is, for given
$\ell$, there are $2\ell+1$ mode frequencies with ${\rm
  Re}(\omega)>0$, and a complementary set with ${\rm Re}(\omega)<0$
that are related by Eq.~(\ref{eqn:freqsymmetry}).  As made clear by
Berti {\em et al}\cite{berticardosowill-2006}, the existence of a
second set of modes satisfying the symmetry of
Eq.~(\ref{eqn:freqsymmetry}) is necessary to asymptotically represent
both polarizations of gravitational waves in terms of QNMs.

The situation is somewhat different for the TTMs.  There are no TTMs
where the ${\rm Re}(\bar\omega)>0$ and $m<0$.  There are TTMs with
$m<0$, but these all have ${\rm Re}(\bar\omega)<0$.  They are
solutions from the complementary set of modes satisfying
Eq.~(\ref{eqn:freqsymmetry}).  The $m=0$ modes also obey this
relationship, so there are a pair of $m=0$ solutions, the $m=0_0$ and
$0_2$ segments are degenerate with their complementary counterparts,
but the ${\rm Re}(\bar\omega)$ of the $m=0_1$ segments have opposite
signs.

Including the $m=0$ case, these {\em known} polynomial modes have only
$\ell+1$ values of $m$ in each of these complementary sets, rather
than the expected $2\ell+1$ values of $m$.  Is it possible that the
remaining $\ell$ values of $m$ that we might expect to find are
actually confluent-Heun functions (general infinite-series solutions)
rather than polynomials?  As mentioned in Sec.~\ref{sec:methods}, the
condition for the existence of polynomial solutions is equivalent to
the vanishing of the magnitude squared of the Starobinsky constant, so
the polynomial solutions are both algebraically special and TTMs.
Both Wald\cite{wald-1973} and Chandrasekhar\cite{chandra-1984} have
shown that the vanishing of the Starobinsky constant is the necessary
and sufficient conditions for algebraically-special solutions.  If all
TTMs are algebraically special, then there can be no TTMs in the form
of confluent-Heun functions.  While it seems that the
algebraically-special solutions should encompass all of the TTMs, we
would like to verify that some subtlety associated with the inner
boundary condition doesn't allow for the existence of confluent-Heun
function TTMs.  We can do this by explicitly searching for them, and
have done so.  An extensive search for confluent-Heun function
solutions for the TTMs\footnote{\label{FN:search}To search for TTMs
  that are confluent-Heun functions, we examined contour plots of the
  real and imaginary parts of the continued fraction function whose
  roots designate the TTMs.  We examined the first 3 $\ell$ values for
  $m=0,\ \pm1,\ \pm2$, and $\pm3$.  For each case, we considered
  $\bar{a}=0.2,\ 0.4,\ 0.6$, and $0.8$.  Finally, we looked for roots
  in the range $-5\le{\rm Re}(\bar\omega)\le25$ and $-30\le{\rm
    Im}(\bar\omega)\le0$.  All candidate roots were further examined
  and found to be polynomial modes.} found no evidence for any such
solutions.  {\em Interestingly, we did find previously unknown
  polynomial TTM (algebraically-special) solutions, all of which start
  at complex infinity in the limit that $\bar{a}\to0$}.  These new
solutions were discovered after this paper was submitted, while we
were verifying the absence of confluent-Heun function TTMs using the
more methodical search described in footnote~\ref{FN:search}.  We are
currently exploring the details of these new TTMs, and will report on
them in a future paper.

\section{Asymptotic behavior of the $m=0$ modes}
\label{sec:asymptotic_behavior}

\begin{figure}
\includegraphics[width=\linewidth,clip]{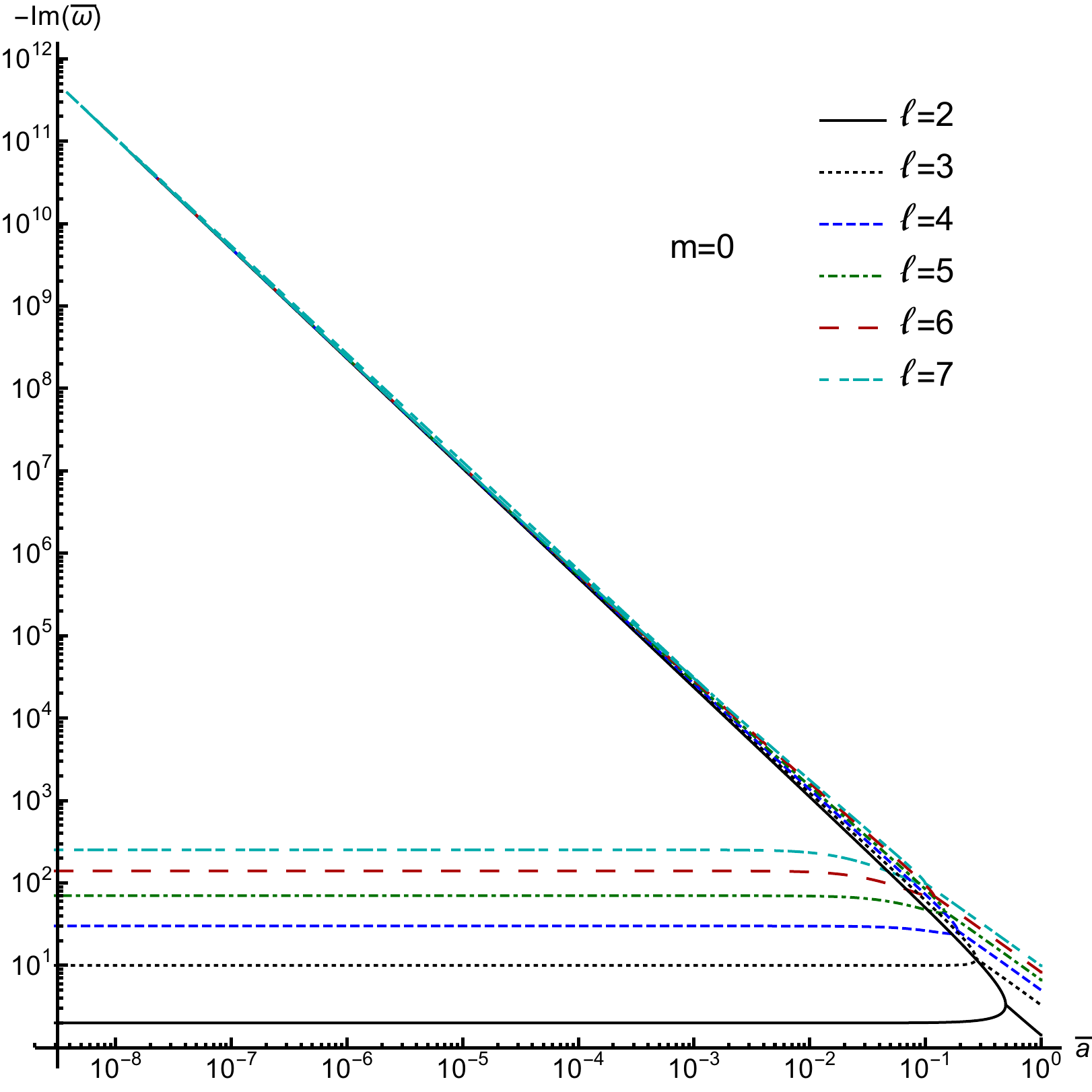}
\caption{\label{fig:TTMloglogm0} $\log$--$\log$ plot of
  $\bar\omega$ for $m=0$ and $\ell=2\mbox{--}7$, clearly showing the
  asymptotic behavior of the $m=0_2$ branches.}
\end{figure}

We will refer to the $m=0_2$ branch of a given $m=0$ TTM sequence as
the {\em asymptotic} branch because we find that
$\lim_{\bar{a}\to0}{\bar\omega(\bar{a})}=-i\infty$ and because it
involves the asymptotic behavior of the separation constant
$\scA[]{s}{\ell m}{\bar{a}\bar\omega}$.  As mentioned above, to our
knowledge, this asymptotic branch had gone unnoticed prior to the work
of Ref.\cite{cook-zalutskiy-2016b}.  Interestingly, a somewhat similar
behavior has recently been noticed in the mode spectrum of rotating
black strings\cite{Mamani-etal-2018}.  Figure~\ref{fig:TTMloglogm0}
presents a $\log$-$\log$ plot of ${\rm
  Im}\left(\bar\omega(\bar{a})\right)$ clearly showing the asymptotic
behavior of the $m=0_0$ branches.  The asymptotic branches have
power-law behavior and appear to be independent of $\ell$ at leading
order.

Simple fitting of the slope shows that the leading order asymptotic
behavior is $\bar\omega\propto a^{-4/3}$.  We could also fit for the
proportionality constant, but we can do better by making use of the
magnitude squared of the Starobinsky constant $|\mathcal{Q}|^2$ as
given in Eq.~(\ref{eq:Starobinsky_const}).  However, to proceed along
this path, we also need to know the asymptotic behavior of the
separation constant $\scA[]{s}{\ell m}{\bar{a}\bar\omega}$.

For mode frequencies on the NIA, the oblateness parameter
$c=a\omega=\bar{a}\bar\omega$ of the angular Teukolsky equation,
Eq.~(\ref{eqn:swSF_DiffEqn}), is purely imaginary.  In a more general
context, the angular Teukolsky equation is the equation for the
spin-weighted spheroidal function defining the $x=\cos\theta$ angular
dependence of the spin-weighted spheroidal harmonics (SWSHs).  For
spin-weight $s=0$, the SWSHs are the spheroidal harmonics where real
values for the oblateness parameter $c$ correspond to an {\em oblate}
spheroidal geometry, while imaginary values of $c$ correspond to {\em
  prolate} spheroidal geometry.  This terminology extends to the
SWSHs.  While general solutions for SWSHs and their associated
eigenvalues $\scA[]{s}{\ell m}{c}$ must be determined numerically,
quite a bit is known about them in certain
limits\cite{berticardosocasals-2006}.  Unfortunately, for the
asymptotic prolate case relevant to us, relatively little is know.
For general values of $\ell$, $m$, and $s$, the asymptotic value of
$\scA[]{s}{\ell m}{c}$ is know only to leading order in $c$, and takes
the form\cite{berticardosocasals-2006}
\begin{equation}\label{eqn:asympt_prolate_0}
  \scA[]{s}{\ell m}{c} = (2L+1)|c| + \mathcal{O}(|c|^0), \qquad|c|\to\infty,
\end{equation}
where
\begin{equation}\label{eqn:Ldef}
  L=\ell - \max(|m|,|s|).
\end{equation}

Fortunately, the known leading order asymptotic behavior of the
separation constant in the prolate case is sufficient to allow us to
determine the asymptotic behavior of the $\bar\omega$.  If we take
$\bar\omega\to-iC_1a^{-4/3}$ and $\scA[]{s}{\ell
  0}{\bar{a}\bar\omega}\to (2L+1)C_1a^{-1/3}$ in
Eq.~(\ref{eq:Starobinsky_const}), and make a power series expansion in
powers of $a$, we find that the vanishing of $|\mathcal{Q}|^2$ yields
\begin{equation}\label{eqn:Starexp1}
  C_1^2\frac{C_1^6-144}{a^{8/3}} - 4C_1^7\frac{2L+1}{a^{7/3}}
  + \mathcal{O}(a^{-7/3})=0.
\end{equation}
We emphasize that this expression is obtained for both $s=\pm2$.  The
vanishing of the leading-order coefficient yields $C_1=12^{1/3}$,
independent of $\ell$ as expected from Fig.~\ref{fig:TTMloglogm0}.
The coefficient of the next term depends on $\ell$, through $L$ from
the leading order behavior of $\scA[]{s}{\ell 0}{c}$.  Vanishing of the
coefficient of the second term depends on higher order behavior of
$\bar\omega$.  For this term to contribute at
$\mathcal{O}(\bar{a}^{-7/3})$ in $|\mathcal{Q}|^2$, the next term in
the expansion of $\bar\omega$ must be $\mathcal{O}(\bar{a}^{-1})$.  If
we repeat the procedure above with
$\bar\omega\to-i(C_1\bar{a}^{-4/3}+C_2\bar{a}^{-1})$, we find that the
vanishing of $|\mathcal{Q}|^2$ yields
\begin{eqnarray}\label{eqn:Starexp2}
  C_1^2\frac{C_1^6-144}{\bar{a}^{8/3}}
  - 4C_1\frac{C_1^6(2L+1)+C_2(72-2C_1^6)}{\bar{a}^{7/3}}
  \nonumber \\ + \mathcal{O}(\bar{a}^{-2})=0.
\end{eqnarray}
Again, this expression is obtained for both $s=\pm2$.  The second term
allows us to determine
\begin{equation}
  C_2=(2L+1)\frac{C_1^6}{2C_1^2-72}.
\end{equation}
Thus, by using $|\mathcal{Q}|^2=0$, knowing only the leading order
term in the asymptotic expansion for $\scA[]{s}{\ell m}{c}$, and
assuming the leading order asymptotic falloff is $\bar{a}^{-4/3}$, we
can determine the first two terms of the asymptotic expansion for
$\bar\omega$ to be
\begin{equation}
  \bar\omega(\bar{a}) = -i\left(\frac{12^{1/3}}{\bar{a}^{4/3}}
  + \frac23\frac{(2\ell-3)}{\bar{a}}\right) + \mathcal{O}(\bar{a}^{-2/3}),
\end{equation}
and we have chosen to write this expression in terms of $\ell$, instead
of $L$, because it is only valid for $m=0$ and $s=\pm2$.

To extend this expansion for $\bar\omega$ further by this method
requires that we know the asymptotic form for $\scA[]{s}{\ell m}{c}$
to higher order.  While this is not known analytically, we can fit
our TTM sequences to determine the unknown terms.  We will use general
expansions of the following forms
\begin{eqnarray}
\label{eqn:omegaexp}
  \bar\omega &=& -i\biggl[\frac{C_1}{\bar{a}^{4/3}} + \frac{C_2}{\bar{a}}
  + \frac{C_{\pm3}}{\bar{a}^{2/3}} + \frac{C_4}{\bar{a}^{1/3}}
  + C_5 \nonumber \\ &&\hspace{1in}\mbox{}
  + C_6\bar{a}^{1/3} + \cdots\biggr], \\
\label{eqn:Aexp}
  \scA[]{\pm2}{\ell 0}{c} &=& (2L+1)|c| + A_{\pm1} + \frac{A_2}{|c|}
  + \frac{A_3}{|c|^2} \nonumber\\ &&\hspace{1in}\mbox{}
  + \frac{A_4}{|c|^3} + \cdots.
\end{eqnarray}
The need for the $\pm$ notation on $C_{\pm3}$ and $A_{\pm1}$ will be
explained in detail below.  The coefficients $C_1$ and $C_2$ have
already been fixed by the vanishing of the first two terms in the
power series expansion of $|\mathcal{Q}|^2=0$.  The third term, after
replacing $C_1$ and $C_2$ yields, for $s=2$
\begin{equation}\label{eqn:c3p_exp}
  C_{+3} = \frac{4L(L+1)-29+6A_{+1}}{9\times12^{1/3}},
\end{equation}
and for $s=-2$
\begin{equation}\label{eqn:c3m_exp}
  C_{\minus3} = \frac{4L(L+1)-53+6A_{\minus1}}{9\times12^{1/3}}.
\end{equation}
So, the $C_{\pm3}$ coefficients can be determined once the $A_{\pm1}$
coefficients are known.

We can find the $A_{\pm1}$ coefficient numerically in 2 independent
ways.  First, we can fit our numerical data for $\bar\omega(\bar{a})$
against Eq.~(\ref{eqn:omegaexp}).  With known values for $C_1$ and
$C_2$, and using Eqs.~(\ref{eqn:c3p_exp}) and (\ref{eqn:c3m_exp}), we
can use least-squares fitting to extract numerical values for
$A_{\pm1}$ from our data for $\bar\omega(\bar{a})$ for $2\le\ell\le7$.
We can then find fitting functions for the extracted values of
$A_{\pm1}(\ell)$.  From the $A_{+1}$ data ($s=2$) we find
\begin{equation}\label{eqn:A1p}
  A_{+1}= -\left(\frac12L(L+1) -\frac54\right).
\end{equation}
From the $A_{-1}$ data ($s=-2$) we find
\begin{equation}\label{eqn:A1m}
  A_{\minus1}= -\left(\frac12L(L+1) -\frac{21}4\right).
\end{equation}
Inserting these into Eqs.~(\ref{eqn:c3p_exp}) and (\ref{eqn:c3m_exp}),
we find the shared result
\begin{equation}\label{eqn:c3_exp}
  C_{\pm3} = \frac{2L(L+1)-43}{18\times12^{1/3}}.
\end{equation}

Even though the TTM${}_{\rm R}$ and TTM${}_{\rm L}$ modes share the
same frequency spectrum, the $C_{\pm3}$ and $A_{\pm1}$ notation is
necessary because $\scA{-s}{\ell{m}}{c}=\scA{s}{\ell{m}}{c}+2s$.  This
difference manifests itself only in the constant term in the expansion
for $\scA[]{s}{\ell m}{c}$.  We can express the $s$ dependence
explicitly in $A_{\pm1}$.  Using the additional information from the
asymptotic expansion for $\scA[]{0}{\ell m}{c}$ (see
Ref.\cite{berticardosocasals-2006}), we get
\begin{equation}\label{eqn:A1_gen}
  A_{\pm1}= -\left(\frac12L(L+1) +\frac34 -s(s-1)\right).
\end{equation}

Following a similar procedure, we find we have enough numerical
precision to accurately extract the $A_2$ and $A_3$ coefficients,
allowing us to obtain the values for $C_4$ and $C_5$.  Our final
result for the asymptotic form for the angular separation constant
is
\begin{eqnarray}\label{eqn:Asympt_A}
  \scA[]{\pm2}{\ell 0}{c} &=& (2L+1)|c|
  - \left(\frac12L(L+1) +\frac34 -s(s-1)\right) \nonumber \\
  &&\mbox{} - \frac{(2L+1)(L(L+1)+3-16s^2)}{2^4|c|} \\
  &&\mbox{} - \frac{5(L^4+2L^3+7L+3)-2s^2(91L^2+96L-80)}{2^6|c|^2} \nonumber \\
  &&\mbox{} + \mathcal{O}(|c|^{-3}).\nonumber
\end{eqnarray}
We strongly emphasize that, even though this expression is written in
a very general way, it is only tested for $m=0$ and $s=\pm2$ or $s=0$.
In particular, the terms involving $s^2$ could be any even power of
$s$ with an appropriate prefactor.  Finally, as the main result from
this work, we find that the asymptotic expansion of the $m=0$ TTM
frequencies is
\begin{eqnarray}\label{eqn:Asympt_omega}
  \bar\omega(\bar{a}) &=& -i\biggl[\frac{12^{1/3}}{\bar{a}^{4/3}}
    + \frac23\frac{(2\ell-3)}{\bar{a}}
    + \frac{2\ell^2 - 6\ell - 39}{18\times12^{1/3}\bar{a}^{2/3}}
    \nonumber \\ &&\mbox{}
    - \frac{(2\ell-3)(19\ell^2 -57\ell + 2187)}{1296\times18^{1/3}\bar{a}^{1/3}}
    \nonumber \\ &&\mbox{}
    + \frac{\ell^4 - 6\ell^3 + 27\ell - 45}{384}\biggr]
    \nonumber \\ &&\mbox{}
  + \mathcal{O}(\bar{a}^{1/3})
\end{eqnarray}

\begin{figure}
\includegraphics[width=\linewidth,clip]{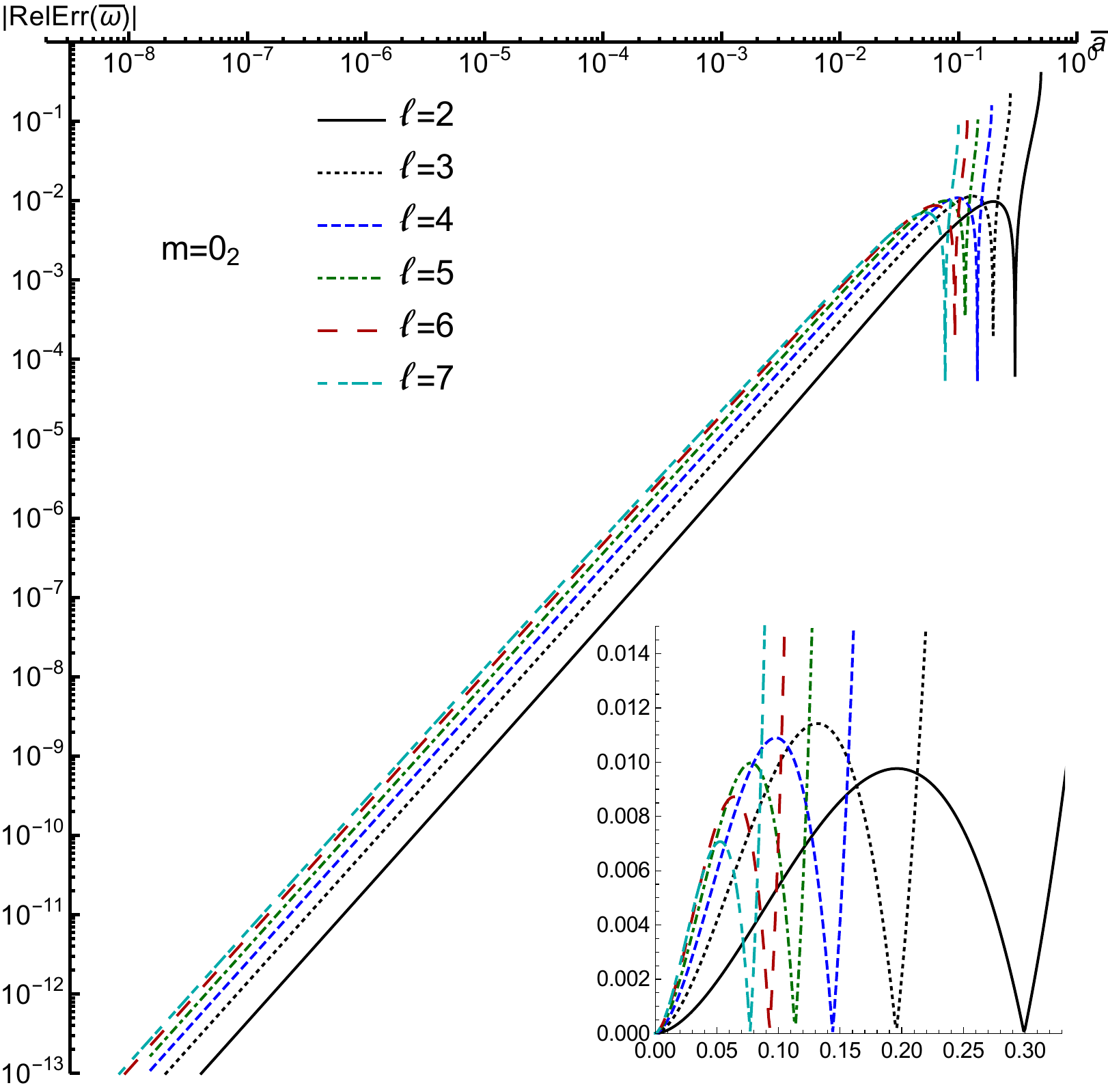}
\caption{\label{fig:TTMloglogomegaerr} $\log$--$\log$ plot of the
  relative error in the asymptotic formula for $\bar\omega$ given in
  Eq.~(\ref{eqn:Asympt_omega}).  The error is computed relative to the
  numerical data for $\bar\omega$ for $m=0$ and $\ell=2\mbox{--}7$.
  As expected, the relative error behaves as $\bar{a}^{5/3}$ in the
  asymptotic regime.  The inset show a standard plot, emphasizing
  the region where the relative errors are the largest.}
\end{figure}

To quantify the accuracy of our expressions for $\bar\omega(\bar{a})$
and $\scA[]{\pm2}{\ell 0}{c}$, we can examine the relative error of
these expressions in comparison with our numerical data.
Figure~\ref{fig:TTMloglogomegaerr} shows the relative error in
Eq.~(\ref{eqn:Asympt_omega}) for $\bar\omega$.  Specifically, we take
the difference between Eq.~(\ref{eqn:Asympt_omega}) and the numerical
data for $\bar\omega$ at the same value of $\bar{a}$, and divide this
by the value of the numerical data.  The absolute value of this
quantity is plotted verses $\bar{a}$.  Clearly, the relative error is
quite small in the asymptotic regime where $\bar{a}\to0$.  Since the
first unknown term in Eq.~(\ref{eqn:Asympt_omega}) goes as
$\bar{a}^{1/3}$ and the leading order term goes as $\bar{a}^{-4/3}$,
we expect the relative error to behave as $\bar{a}^{5/3}$ in the
asymptotic regime.  This is precisely what is found.  Remarkably,
Eq.~(\ref{eqn:Asympt_omega}) is a very good approximation until
reasonably close to the critical value of $\bar{a}$ at which the
$m=0_{0,1,2}$ branches meet.  Prior to this point on the $m=0_2$
branch, we see a ``zero crossing'' in the relative error for each
$\ell$, and up until this point, the relative error is always less
than $1.2\%$.

\begin{figure}\vspace{0.1in}
\includegraphics[width=\linewidth,clip]{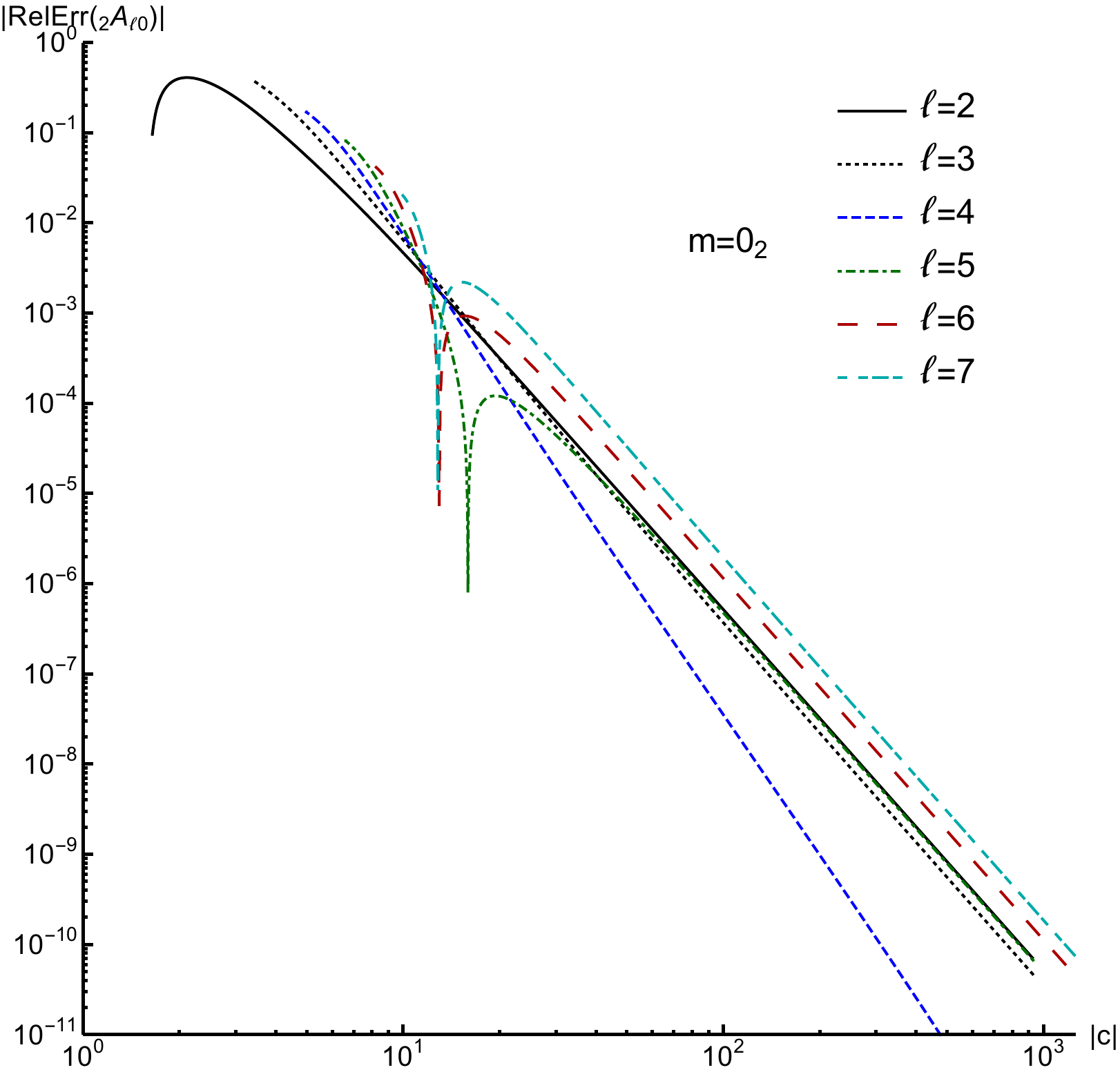}
\caption{\label{fig:TTMloglogAerr} $\log$--$\log$ plot of the relative
  error in the asymptotic formula for $\scA{2}{\ell0}{c}$ given in
  Eq.~(\ref{eqn:Asympt_A}).  The error is computed relative to the
  numerical data for $\scA{2}{\ell0}{c}$ for $m=0$ and $\ell=2\mbox{--}7$.
  As expected, the relative error behaves as $|c|^{-4}$ in the
  asymptotic regime.}
\end{figure}

Figure~\ref{fig:TTMloglogAerr} displays a similar plot showing the
relative error in Eq.~(\ref{eqn:Asympt_A}) for $\scA[]{\pm2}{\ell
  0}{c}$. Specifically, we take the difference between
Eq.~(\ref{eqn:Asympt_A}) and the numerical data for$\scA[]{2}{\ell
  0}{c}$ at the same value of $c=\bar{a}\bar\omega$, and divide this
by the value of the numerical data.  The absolute value of this
quantity is plotted verses $|c|$, and we remind the reader that
$c$ is purely imaginary.  Again, we see that the relative error
is quite small in the asymptotic regime where $|c|\to\infty$.  Since
the first unknown term in Eq.~(\ref{eqn:Asympt_A}) goes as $|c|^{-3}$
and the leading order term goes as $|c|$, we expect the relative
error to behave as $|c|^{-4}$.  This is found for all values of
$\ell$ except for $\ell=4$.  In this case the relative error
falls off even faster, $\approx|c|^{-21/5}$.  This is most likely
due to the $\ell=4$ sequence approaching a ``zero crossing'' which
we see clearly happening for $\ell=5$--$7$.

\begin{figure}
\includegraphics[width=\linewidth,clip]{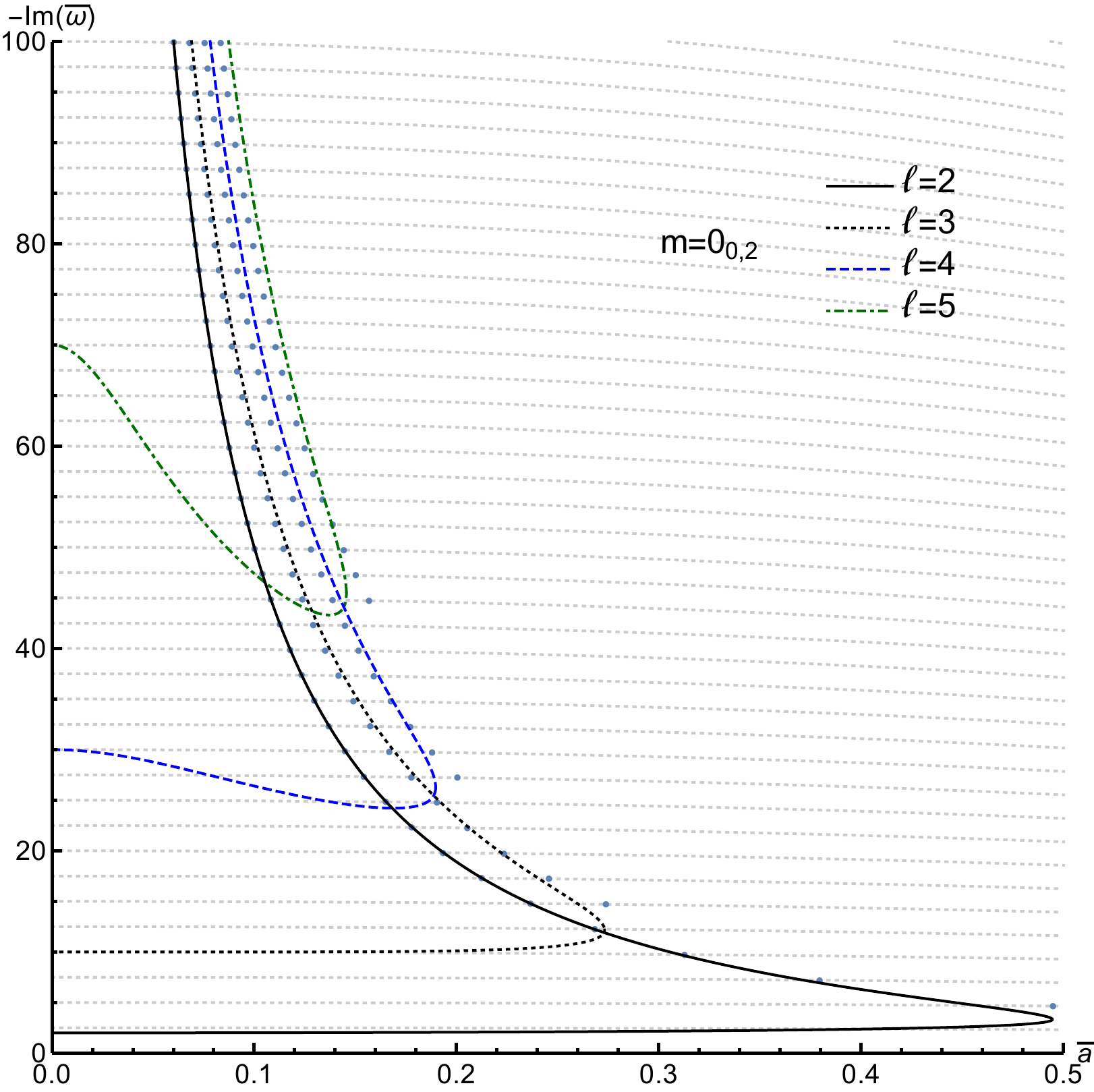}
\caption{\label{fig:TTManomalous} The locations of non-generic modes
  are illustrated in this plot. ${\rm Im}(\bar\omega)$ is plotted as a
  function of $\bar{a}$.  The numerical data for the TTMs with $m=0_0$
  and $\ell=2$--$5$ are plotted as denoted in the figure legend.  The
  $m=0_1$ branches, where $\bar\omega$ contains a non-vanishing real
  component are omitted.  The nearly horizontal gray dotted lines show
  $\bar\omega_+$ as functions of $\bar{a}$ for the values
  $N_+=10,\ 20,\ 30,\cdots$.  The intersection of these gray dotted
  lines with lines denoting the TTMs are locations where the modes are
  anomalous.  The individual dots denote the estimated location of the
  anomalous modes based on Eq.~(\ref{eqn:omegaplus}).}
\end{figure}

\begin{figure}\vspace{0.1in}
\includegraphics[width=\linewidth,clip]{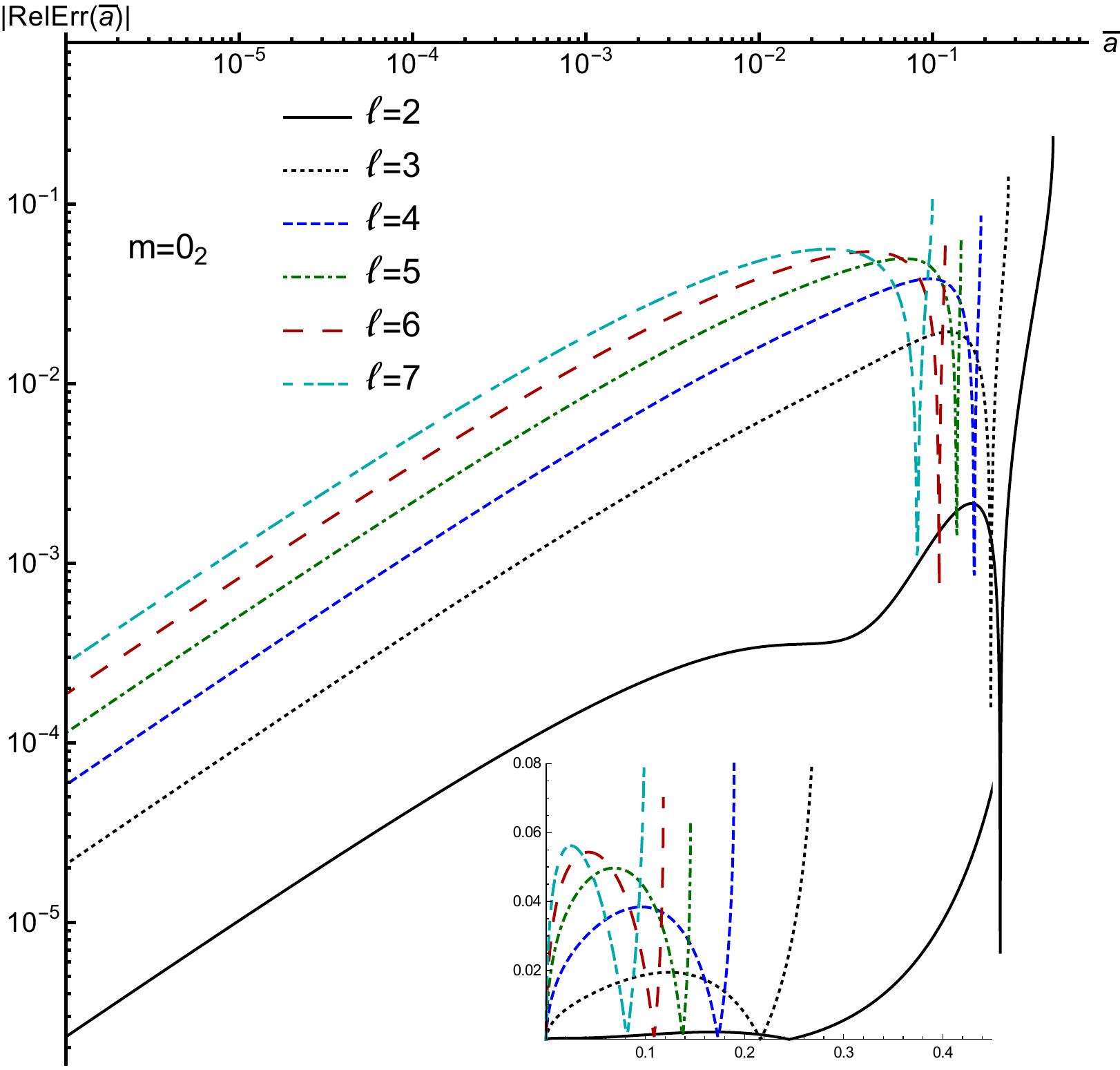}
\caption{\label{fig:TTManomaerror} $\log$--$\log$ plot of the relative
  error in the estimated value of $\bar{a}$ for the non-generic modes as
  a function of $\bar{a}$.  The inset shows a standard plot, emphasizing
  the region where the errors are the largest.}
\end{figure}

\section{Summary and Discussion}
\label{sec:summary}

The $m=0_0$ and $0_2$ branches of the various $m=0$ TTMs represent
very special modes for which the mode frequency is purely imaginary,
thus representing modes that are simply decaying with no oscillation.
These are the only gravitational modes of the Kerr geometry known to
have purely imaginary frequencies, and the $m=0_2$ branches were
unknown before being noticed in Ref.~\cite{cook-zalutskiy-2016b},
where the framework to fully understand the behavior of these modes
was worked out.

For the most-part, these modes are TTMs with the TTM${}_{\rm L}$s and
the TTM${}_{\rm R}$s sharing the same mode frequencies.  However, for
each $\ell$, there are a countably-infinite number of frequencies
where the modes are either ``anomalous'' or ``miraculous'' in the
language of Ref.~\cite{cook-zalutskiy-2016b}.  At these special
frequencies, the modes that would normally be just TTM${}_{\rm L}$s
are also simultaneously QNMs.  Furthermore the modes that would
normally be just TTM${}_{\rm R}$s are neither TTMs nor QNMs.

This non-generic behavior occurs when the roots of the indicial
equation differ by an integer for the Frobenius solution local to the
event horizon.  The condition for this is
\begin{equation}\label{eqn:omegaplus}
  \bar\omega=\bar\omega_+\equiv
  \frac{\bar{a}m-iN_+\sqrt{1-\bar{a}^2}}{2(1+\sqrt{1-\bar{a}^2})},
\end{equation}
where $N_+$ is an integer when $s$ is an integer, and it is a half-odd
integer when $s$ is a half-odd integer.  Two possible non-generic
behaviors can occur.  In the simplest case, only one series solution
can exist, and it corresponds to the local Frobenius solution with
the largest exponent.  The second solution will include a logarithmic
term.  This is the {\em anomalous} case.  In certain circumstances,
the coefficient multiplying the logarithmic term can vanish.  This is
the {\em miraculous} case.

For gravitational modes of Kerr, when $s=-2$, the $m=0_0$ and $0_2$
branches are normally TTM${}_{\rm L}$s.  But when the mode frequencies
satisfy Eq.~(\ref{eqn:omegaplus}) for some integer value of $N_+$,
then the solutions are {\em anomalous}.  The outgoing mode in this
case has exactly the same local behavior as the incoming
mode.\footnote{This is due to strong scattering of the outgoing mode
  off of the potential tail.  The scattering is so strong that it
  overwhelms the normally dominant behavior of the outgoing mode.  See
  Sec.~IV.D of Ref.~\cite{cook-zalutskiy-2016b} for details.}
Therefore, the $s=-2$ modes with frequencies satisfying
Eq.~(\ref{eqn:omegaplus}) are simultaneously TTM${}_{\rm L}$s and
QNMs.  But, for the $s=+2$ modes which are normally TTM${}_{\rm R}$s,
when the mode frequencies satisfy Eq.~(\ref{eqn:omegaplus}) for some
integer value of $N_+$, then the solutions are {\em miraculous}.
In this case, the behavior local to the event horizon is a linear
combination of incoming and outgoing modes, violating the
conditions for the modes to be TTM${}_{\rm R}$s.

The locations where these non-generic modes occur is illustrated in
Fig.~\ref{fig:TTManomalous}.  For clarity, only the $m=0_0$ and $0_2$
mode sequences for $\ell=2$--$5$ from the numerical data are included.
In addition, $\omega_+$ from Eq.~(\ref{eqn:omegaplus}) is plotted as a
function of $\bar{a}$.  Each nearly horizontal dotted-gray line
represents a specific value of $N_+$.  Again for clarity, only the
values of $N_+=10,\ 20,\ 30,\cdots$ are plotted.  Non-generic modes
occur whenever the dotted-gray lines cross one of the sequences of
numerical modes.\footnote{These crossings are precisely what was
  plotted as black dots in Fig.~20 of
  Ref.~\cite{cook-zalutskiy-2016b}.}  We emphasize that only $1$ in
$10$ of the crossing is actually shown in Fig.~\ref{fig:TTManomalous}.

We can find the approximate value of $\bar{a}$ for each of the
crossings of the $m=0_2$ branches by equating
Eqs.~(\ref{eqn:Asympt_omega}) and (\ref{eqn:omegaplus}).  To leading
order, we find $\bar{a}_+\approx 4\cdot3^{1/4}N^{-3/4}$.  This is
independent of $\ell$, and while it gives a good approximation to
$\bar\omega$ at each crossing, it is not a good approximation for
the value of $\bar{a}$ itself.

If we include all the terms in Eq.~(\ref{eqn:Asympt_omega}), but only
the leading order term in the expansion of Eq.~(\ref{eqn:omegaplus}),
then we have a system that is consistent to ${\mathcal
  O}(\bar{a}^{1/3})$.  If we take
$\bar{a}_+=4\cdot3^{1/4}N^{-3/4}(1+\alpha)$ and assume $\alpha$ is small,
expanding to linear order in $\alpha$ finally yields
\begin{widetext}
\begin{equation}\label{eqn:acrossing}
  \bar{a}_+\approx \frac{3456\cdot3^{1/4}N_+^{1/2} + 1152\,(2\ell-3)N_+^{1/4}
    - 16\cdot3^{3/4}(14\ell^2-42\ell+75)}{864\,N_+^{5/4}
    - 48\cdot3^{3/4}(2\ell-3)N_+ + 8\cdot3^{1/2}(14\ell^2 - 42\ell+75)N_+^{3/4}
    - 9\cdot3^{1/4}(2\ell-3)(7\ell^2-21\ell-25)N_+^{1/2}}.
\end{equation}
\end{widetext}
In Fig.~\ref{fig:TTManomalous}, the small dots represent the
approximation for the location of the crossings using
Eqs.~(\ref{eqn:acrossing}) and (\ref{eqn:omegaplus}).  The
approximation is clearly excellent for $\ell=2$, with increasing error
as $\ell$ increases.  Figure~\ref{fig:TTManomaerror} displays the
relative error of the approximate location of the crossings.  It is
obtained by treating $N_+$ as a continuum variable and obtaining its
value for each point in the numerical data along the asymptotic
branches.  Using $\ell$, $\bar{a}$, and the so-obtained value of $N_+$
from each numerical sequence, we can compute the error relative to the
actual value of $\bar{a}$.  In this way, Fig.~\ref{fig:TTManomaerror}
shows the relative error inherent in Eq.~(\ref{eqn:acrossing}).

These asymptotic modes are intriguing, and while we have been able to
elucidate their properties with significant precision, their physical
importance is less clear.  All of the modes of concern have
frequencies that lie on the NIA, and are restricted to $m=0$.  Because
the frequencies are purely imaginary, the individual modes represent
pure damping states.  Furthermore, these are all highly damped states.
For comparison, the fastest damping fundamental $\ell=2$ QNM is more
than 20 times smaller than the slowest damping mode on the NIA with
$\bar\omega=-2i$.

The $s=-2$ gravitational modes with frequencies on the NIA are all
TTM${}_{\rm L}$s, but at certain values of $\bar{a}$, they are also
simultaneously QNMs.  At these same values of $\bar{a}$, the $s=2$
TTM${}_{\rm R}$s with modes frequencies on the NIA do not satisfy the
boundary conditions for TTM${}_{\rm R}$s.  Given this behavior, we are
tempted to think of the set of modes which are simultaneously
TTM${}_{\rm L}$ and QNM as somehow, in the asymptotic limit, hinting
at a discreteness in the allowed values of $\bar{a}$.  Certainly,
there is a ``correspondence principle'' like behavior as $\bar{a}\to0$
where the spacing between adjacent values of $\bar{a}$ becomes
vanishingly small with $\Delta\bar{a}\sim\bar{a}^{7/3}$.

However, our consideration of this very regular behavior in these
particular modes on the NIA is ignoring another set of QNMs that exist
on the NIA and satisfy $\bar\omega_\minus=-iN_\minus/4$ (see
Ref.\cite{cook-zalutskiy-2016b}).  These modes do not exist for all
allowed values of $N_\minus$, and the determination of the value of
$\bar{a}$ at which these QNM do exist is less easily analyzed.
Furthermore, these special modes that are simultaneously QNM and
TTM${}_{\rm L}$ are not isolated modes but rather represent
termination points of sequences of modes that otherwise do not have
frequencies on the NIA.

In conclusion, we cannot yet say if these modes carry any important
physical significance.  Never-the-less, they are present in the
QNM/TTM spectrum.

\acknowledgments Some computations were performed on the Wake Forest
University DEAC Cluster, a centrally managed resource with support
provided in part by the University.


\begin{thebibliography}{18}%
\makeatletter
\providecommand \@ifxundefined [1]{%
 \@ifx{#1\undefined}
}%
\providecommand \@ifnum [1]{%
 \ifnum #1\expandafter \@firstoftwo
 \else \expandafter \@secondoftwo
 \fi
}%
\providecommand \@ifx [1]{%
 \ifx #1\expandafter \@firstoftwo
 \else \expandafter \@secondoftwo
 \fi
}%
\providecommand \natexlab [1]{#1}%
\providecommand \enquote  [1]{``#1''}%
\providecommand \bibnamefont  [1]{#1}%
\providecommand \bibfnamefont [1]{#1}%
\providecommand \citenamefont [1]{#1}%
\providecommand \href@noop [0]{\@secondoftwo}%
\providecommand \href [0]{\begingroup \@sanitize@url \@href}%
\providecommand \@href[1]{\@@startlink{#1}\@@href}%
\providecommand \@@href[1]{\endgroup#1\@@endlink}%
\providecommand \@sanitize@url [0]{\catcode `\\12\catcode `\$12\catcode
  `\&12\catcode `\#12\catcode `\^12\catcode `\_12\catcode `\%12\relax}%
\providecommand \@@startlink[1]{}%
\providecommand \@@endlink[0]{}%
\providecommand \url  [0]{\begingroup\@sanitize@url \@url }%
\providecommand \@url [1]{\endgroup\@href {#1}{\urlprefix }}%
\providecommand \urlprefix  [0]{URL }%
\providecommand \Eprint [0]{\href }%
\providecommand \doibase [0]{http://dx.doi.org/}%
\providecommand \selectlanguage [0]{\@gobble}%
\providecommand \bibinfo  [0]{\@secondoftwo}%
\providecommand \bibfield  [0]{\@secondoftwo}%
\providecommand \translation [1]{[#1]}%
\providecommand \BibitemOpen [0]{}%
\providecommand \bibitemStop [0]{}%
\providecommand \bibitemNoStop [0]{.\EOS\space}%
\providecommand \EOS [0]{\spacefactor3000\relax}%
\providecommand \BibitemShut  [1]{\csname bibitem#1\endcsname}%
\let\auto@bib@innerbib\@empty
%</preamble>
\bibitem [{\citenamefont {Kerr}(1963)}]{kerr-1963}%
  \BibitemOpen
  \bibfield  {author} {\bibinfo {author} {\bibfnamefont {R.~P.}\ \bibnamefont
  {Kerr}},\ }\href {\doibase 10.1103/PhysRevLett.11.237} {\bibfield  {journal}
  {\bibinfo  {journal} {Phys. Rev. Lett.}\ }\textbf {\bibinfo {volume} {11}},\
  \bibinfo {pages} {237} (\bibinfo {year} {1963})}\BibitemShut {NoStop}%
\bibitem [{\citenamefont {Abbott}\ \emph
  {et~al.}(2016{\natexlab{a}})\citenamefont {Abbott} \emph
  {et~al.}}]{GW150914-2016}%
  \BibitemOpen
  \bibfield  {author} {\bibinfo {author} {\bibfnamefont {B.~P.}\ \bibnamefont
  {Abbott}} \emph {et~al.} (\bibinfo {collaboration} {LIGO Scientific
  Collaboration and Virgo Collaboration}),\ }\href {\doibase
  10.1103/PhysRevLett.116.061102} {\bibfield  {journal} {\bibinfo  {journal}
  {Phys. Rev. Lett.}\ }\textbf {\bibinfo {volume} {116}},\ \bibinfo {pages}
  {061102} (\bibinfo {year} {2016}{\natexlab{a}})}\BibitemShut {NoStop}%
\bibitem [{\citenamefont {Abbott}\ \emph
  {et~al.}(2016{\natexlab{b}})\citenamefont {Abbott} \emph
  {et~al.}}]{GW151226-2016}%
  \BibitemOpen
  \bibfield  {author} {\bibinfo {author} {\bibfnamefont {B.~P.}\ \bibnamefont
  {Abbott}} \emph {et~al.} (\bibinfo {collaboration} {LIGO Scientific
  Collaboration and Virgo Collaboration}),\ }\href {\doibase
  10.1103/PhysRevLett.116.241103} {\bibfield  {journal} {\bibinfo  {journal}
  {Phys. Rev. Lett.}\ }\textbf {\bibinfo {volume} {116}},\ \bibinfo {pages}
  {241103} (\bibinfo {year} {2016}{\natexlab{b}})}\BibitemShut {NoStop}%
\bibitem [{\citenamefont {Abbott}\ \emph
  {et~al.}(2017{\natexlab{a}})\citenamefont {Abbott} \emph
  {et~al.}}]{GW170104-2017}%
  \BibitemOpen
  \bibfield  {author} {\bibinfo {author} {\bibfnamefont {B.~P.}\ \bibnamefont
  {Abbott}} \emph {et~al.} (\bibinfo {collaboration} {LIGO Scientific
  Collaboration and Virgo Collaboration}),\ }\href {\doibase
  10.1103/PhysRevLett.118.221101} {\bibfield  {journal} {\bibinfo  {journal}
  {Phys. Rev. Lett.}\ }\textbf {\bibinfo {volume} {118}},\ \bibinfo {pages}
  {221101} (\bibinfo {year} {2017}{\natexlab{a}})}\BibitemShut {NoStop}%
\bibitem [{\citenamefont {Abbott}\ \emph
  {et~al.}(2017{\natexlab{b}})\citenamefont {Abbott} \emph
  {et~al.}}]{GW170814-2017}%
  \BibitemOpen
  \bibfield  {author} {\bibinfo {author} {\bibfnamefont {B.~P.}\ \bibnamefont
  {Abbott}} \emph {et~al.} (\bibinfo {collaboration} {LIGO Scientific
  Collaboration and Virgo Collaboration}),\ }\href {\doibase
  10.1103/PhysRevLett.119.141101} {\bibfield  {journal} {\bibinfo  {journal}
  {Phys. Rev. Lett.}\ }\textbf {\bibinfo {volume} {119}},\ \bibinfo {pages}
  {141101} (\bibinfo {year} {2017}{\natexlab{b}})}\BibitemShut {NoStop}%
\bibitem [{\citenamefont {Wald}(1973)}]{wald-1973}%
  \BibitemOpen
  \bibfield  {author} {\bibinfo {author} {\bibfnamefont {R.~M.}\ \bibnamefont
  {Wald}},\ }\href {\doibase 10.1063/1.1666203} {\bibfield  {journal} {\bibinfo
   {journal} {J.~Math. Phys. (N.Y.)}\ }\textbf {\bibinfo {volume} {14}},\
  \bibinfo {pages} {1453} (\bibinfo {year} {1973})}\BibitemShut {NoStop}%
\bibitem [{\citenamefont {Chandrasekhar}(1984)}]{chandra-1984}%
  \BibitemOpen
  \bibfield  {author} {\bibinfo {author} {\bibfnamefont {S.}~\bibnamefont
  {Chandrasekhar}},\ }\href {\doibase 10.1098/rspa.1984.0021} {\bibfield
  {journal} {\bibinfo  {journal} {Proc. R. Soc. A}\ }\textbf {\bibinfo {volume}
  {392}},\ \bibinfo {pages} {1} (\bibinfo {year} {1984})}\BibitemShut {NoStop}%
\bibitem [{\citenamefont {Onozawa}(1997)}]{onozawa-1997}%
  \BibitemOpen
  \bibfield  {author} {\bibinfo {author} {\bibfnamefont {H.}~\bibnamefont
  {Onozawa}},\ }\href {\doibase 10.1103/PhysRevD.55.3593} {\bibfield  {journal}
  {\bibinfo  {journal} {Phys. Rev. D}\ }\textbf {\bibinfo {volume} {55}},\
  \bibinfo {pages} {3593} (\bibinfo {year} {1997})}\BibitemShut {NoStop}%
\bibitem [{\citenamefont {Cook}\ and\ \citenamefont
  {Zalutskiy}(2014)}]{cook-zalutskiy-2014}%
  \BibitemOpen
  \bibfield  {author} {\bibinfo {author} {\bibfnamefont {G.~B.}\ \bibnamefont
  {Cook}}\ and\ \bibinfo {author} {\bibfnamefont {M.}~\bibnamefont
  {Zalutskiy}},\ }\href {\doibase 10.1103/PhysRevD.90.124021} {\bibfield
  {journal} {\bibinfo  {journal} {Phys. Rev. D}\ }\textbf {\bibinfo {volume}
  {90}},\ \bibinfo {pages} {124021} (\bibinfo {year} {2014})}\BibitemShut
  {NoStop}%
\bibitem [{\citenamefont {Leaver}(1985)}]{leaver-1985}%
  \BibitemOpen
  \bibfield  {author} {\bibinfo {author} {\bibfnamefont {E.~W.}\ \bibnamefont
  {Leaver}},\ }\href {\doibase 10.1098/rspa.1985.0119} {\bibfield  {journal}
  {\bibinfo  {journal} {Proc. R. Soc. A}\ }\textbf {\bibinfo {volume} {402}},\
  \bibinfo {pages} {285} (\bibinfo {year} {1985})}\BibitemShut {NoStop}%
\bibitem [{\citenamefont {Maassen van~den Brink}(2000)}]{van_den_brink-2000}%
  \BibitemOpen
  \bibfield  {author} {\bibinfo {author} {\bibfnamefont {A.}~\bibnamefont
  {Maassen van~den Brink}},\ }\href {\doibase 10.1103/PhysRevD.62.064009}
  {\bibfield  {journal} {\bibinfo  {journal} {Phys. Rev. D}\ }\textbf {\bibinfo
  {volume} {62}},\ \bibinfo {pages} {064009} (\bibinfo {year}
  {2000})}\BibitemShut {NoStop}%
\bibitem [{\citenamefont {Cook}\ and\ \citenamefont
  {Zalutskiy}(2016{\natexlab{a}})}]{cook-zalutskiy-2016b}%
  \BibitemOpen
  \bibfield  {author} {\bibinfo {author} {\bibfnamefont {G.~B.}\ \bibnamefont
  {Cook}}\ and\ \bibinfo {author} {\bibfnamefont {M.}~\bibnamefont
  {Zalutskiy}},\ }\href {\doibase 10.1103/PhysRevD.94.104074} {\bibfield
  {journal} {\bibinfo  {journal} {Phys. Rev. D}\ }\textbf {\bibinfo {volume}
  {94}},\ \bibinfo {pages} {104074} (\bibinfo {year}
  {2016}{\natexlab{a}})}\BibitemShut {NoStop}%
\bibitem [{\citenamefont {Cook}\ and\ \citenamefont
  {Zalutskiy}(2016{\natexlab{b}})}]{cook-zalutskiy-2016a}%
  \BibitemOpen
  \bibfield  {author} {\bibinfo {author} {\bibfnamefont {G.~B.}\ \bibnamefont
  {Cook}}\ and\ \bibinfo {author} {\bibfnamefont {M.}~\bibnamefont
  {Zalutskiy}},\ }\href {\doibase 10.1088/0264-9381/33/24/245008} {\bibfield
  {journal} {\bibinfo  {journal} {Classical Quantum Gravity}\ }\textbf
  {\bibinfo {volume} {33}},\ \bibinfo {pages} {245008} (\bibinfo {year}
  {2016}{\natexlab{b}})}\BibitemShut {NoStop}%
\bibitem [{\citenamefont {Ronveaux}(1995)}]{Heun-eqn}%
  \BibitemOpen
  \bibinfo {editor} {\bibfnamefont {A.}~\bibnamefont {Ronveaux}},\ ed.,\
  \href@noop {} {\emph {\bibinfo {title} {Heun's Differential Equations}}}\
  (\bibinfo  {publisher} {Oxford University, New York},\ \bibinfo {year}
  {1995})\BibitemShut {NoStop}%
\bibitem [{\citenamefont {Borissov}\ and\ \citenamefont
  {Fiziev}(2010)}]{Fiziev-2009b}%
  \BibitemOpen
  \bibfield  {author} {\bibinfo {author} {\bibfnamefont {R.~S.}\ \bibnamefont
  {Borissov}}\ and\ \bibinfo {author} {\bibfnamefont {P.~P.}\ \bibnamefont
  {Fiziev}},\ }\href@noop {} {\bibfield  {journal} {\bibinfo  {journal} {Bulg.
  J. Phys.}\ }\textbf {\bibinfo {volume} {37}},\ \bibinfo {pages} {65}
  (\bibinfo {year} {2010})}\BibitemShut {NoStop}%
\bibitem [{\citenamefont {Berti}\ \emph
  {et~al.}(2006{\natexlab{a}})\citenamefont {Berti}, \citenamefont {Cardoso},\
  and\ \citenamefont {Will}}]{berticardosowill-2006}%
  \BibitemOpen
  \bibfield  {author} {\bibinfo {author} {\bibfnamefont {E.}~\bibnamefont
  {Berti}}, \bibinfo {author} {\bibfnamefont {V.}~\bibnamefont {Cardoso}}, \
  and\ \bibinfo {author} {\bibfnamefont {C.~M.}\ \bibnamefont {Will}},\ }\href
  {\doibase 10.1103/PhysRevD.73.064030} {\bibfield  {journal} {\bibinfo
  {journal} {Phys. Rev. D}\ }\textbf {\bibinfo {volume} {73}},\ \bibinfo
  {pages} {064030} (\bibinfo {year} {2006}{\natexlab{a}})}\BibitemShut
  {NoStop}%
\bibitem [{\citenamefont {Mamani}\ \emph {et~al.}(2018)\citenamefont {Mamani},
  \citenamefont {Morgan}, \citenamefont {Miranda},\ and\ \citenamefont
  {Zanchin}}]{Mamani-etal-2018}%
  \BibitemOpen
  \bibfield  {author} {\bibinfo {author} {\bibfnamefont {L.~A.~H.}\
  \bibnamefont {Mamani}}, \bibinfo {author} {\bibfnamefont {J.}~\bibnamefont
  {Morgan}}, \bibinfo {author} {\bibfnamefont {A.~S.}\ \bibnamefont {Miranda}},
  \ and\ \bibinfo {author} {\bibfnamefont {V.~T.}\ \bibnamefont {Zanchin}},\
  }\href {\doibase 10.1103/PhysRevD.98.026006} {\bibfield  {journal} {\bibinfo
  {journal} {Phys. Rev. D}\ }\textbf {\bibinfo {volume} {98}},\ \bibinfo
  {pages} {026006} (\bibinfo {year} {2018})}\BibitemShut {NoStop}%
\bibitem [{\citenamefont {Berti}\ \emph
  {et~al.}(2006{\natexlab{b}})\citenamefont {Berti}, \citenamefont {Cardoso},\
  and\ \citenamefont {Casals}}]{berticardosocasals-2006}%
  \BibitemOpen
  \bibfield  {author} {\bibinfo {author} {\bibfnamefont {E.}~\bibnamefont
  {Berti}}, \bibinfo {author} {\bibfnamefont {V.}~\bibnamefont {Cardoso}}, \
  and\ \bibinfo {author} {\bibfnamefont {M.}~\bibnamefont {Casals}},\ }\href
  {\doibase 10.1103/PhysRevD.73.024013} {\bibfield  {journal} {\bibinfo
  {journal} {Phys. Rev. D}\ }\textbf {\bibinfo {volume} {73}},\ \bibinfo
  {pages} {024013} (\bibinfo {year} {2006}{\natexlab{b}})}\BibitemShut
  {NoStop}%
\end{thebibliography}
\end{document}